\documentclass[nofootinbib,amsmath,amssymb,amsfonts,aps,prx,eqsecnum]{revtex4-1}

\pdfoutput=1
\usepackage{hyperref}
\usepackage{amsmath}
\usepackage{mathrsfs}
\usepackage{xcolor,graphicx}
\usepackage[caption=false]{subfig}

\begin{document}

\title{A Holographic Path to the Turbulent Side of Gravity}

\author{Stephen R. Green}
\email{sgreen04@uoguelph.ca}
\thanks{CITA National Fellow}
\affiliation{Department of Physics \\
  University of Guelph \\
  Guelph, Ontario N1G 2W1, Canada}

\author{Federico Carrasco}
\email{fedecarrasco@gmail.com}
\affiliation{FaMAF-UNC, IFEG-CONICET, Ciudad Universitaria, 5000 \\
  Cordoba, Argentina}

\author{Luis Lehner}
\email{llehner@perimeterinstitute.ca}
\affiliation{Perimeter Institute for Theoretical Physics \\
  31 Caroline Street North \\
  Waterloo, Ontario N2L 2Y5, Canada}

\date{\today}

\begin{abstract}
  We study the dynamics of a $2+1$ dimensional relativistic {\em
    viscous} conformal fluid in Minkowski spacetime.
  Such fluid solutions arise as duals, under the ``gravity/fluid
  correspondence'', to $3+1$ dimensional asymptotically anti-de Sitter
  (AAdS) black brane solutions to the Einstein equation.  We examine
  stability properties of shear flows, which correspond to
  hydrodynamic quasinormal modes of the black brane.  We find that,
  for sufficiently high Reynolds number, the solution undergoes an
  inverse turbulent cascade to long wavelength modes.
  We then map this fluid solution, via the gravity/fluid duality, into
  a bulk metric.  This suggests a new and interesting feature of the
  behavior of perturbed AAdS black holes and black branes, which is
  not readily captured by a standard quasinormal mode analysis.
  Namely, for sufficiently large perturbed black objects (with
  long-lived quasinormal modes), nonlinear effects transfer energy
  from short to long wavelength modes via a turbulent cascade within
  the metric perturbation.  As long wavelength modes have slower
  decay, this lengthens the overall lifetime of the perturbation.  We
  also discuss various implications of this behavior, including
  expectations for higher dimensions, and the possibility of predicting
  turbulence in more general gravitational scenarios.


\end{abstract}

\maketitle

\section{Introduction}

The AdS/CFT correspondence~\cite{Maldacena:1997re,Aharony:1999ti}
proposes a remarkable connection between quantum gravity in $d+1$
dimensions and quantum field theory in $d$ dimensions.  In a certain
classical limit, this correspondence can be utilized to link the
behavior of perturbed asymptotically anti-de Sitter (AAdS) black
branes in general relativity to that of viscous conformal fluids on
the AdS boundary, provided the perturbations are of sufficiently long
wavelength~\cite{Baier:2007ix,Bhattacharyya:2008jc,Bhattacharyya:2008mz,VanRaamsdonk:2008fp}.
This limit of the AdS/CFT correspondence is known as the {\em
  gravity/fluid correspondence}.

The gravity/fluid correspondence can also be derived on its own in a
purely classical manner without any appeal to AdS/CFT, as a derivative
expansion within general relativity (see, e.g.,
Ref.~\cite{Bhattacharyya:2008jc}).  This derivation provides an
explicit perturbative mapping between solutions, which can be
exploited to relate gravitational and fluid behavior.  In particular,
interesting known phenomena on one side of the duality should have
counterparts on the other, which can lead to new predictions, or to
new methods of analysis.  This has been used as a means to frame fluid
dynamics questions in terms of gravitational physics. For example,
Ref.~\cite{Oz:2010wz} explored the relation between the Penrose
inequalities -- which predict the onset of naked singularities in
general relativity -- and finite-time blowup of solutions in
hydrodynamics.  In Ref.~\cite{Eling:2010vr}, it was suggested that a
gravity dual could be utilized to understand the complex
phenomenon of fluid turbulence.  In the present work, following the analysis
presented in Ref.~\cite{Carrasco:2012nf}, we follow the opposite
route, namely to study the implications that turbulent
phenomena -- that can arise in fluid dynamics -- have for our
understanding of general relativity.

Turbulence is a ubiquitous property of fluid flows observed in nature
at sufficiently high Reynolds number, $R$. Such behavior has recently
been shown to also arise in {\em inviscid} conformal relativistic
hydrodynamics~\cite{Carrasco:2012nf,Radice:2012pq}.  While the actual
fluid dual to an AAdS black brane has nonzero shear viscosity, this
viscosity is subleading in the black hole temperature, so the
inviscid approximation is valid at sufficiently high
temperature~\cite{Carrasco:2012nf}.  This suggests that there should
be a corresponding regime where long-wavelength black hole
perturbations in asymptotically AdS spacetimes behave in a turbulent
manner.  Furthermore, in $2+1$ dimensions such inviscid conformal
fluids display an inverse cascade of energy to large
scales~\cite{Carrasco:2012nf}, in accord with intuition from
Navier-Stokes fluids~\cite{Kraichnan:1967}.  This ensures that
provided the initial condition falls within the regime of
applicability of the gravity/fluid correspondence (i.e., sufficiently
long wavelength perturbations), so should its time evolution, and
therefore there should exist a black brane that behaves in a dual
manner.  The intuition described here has been borne out in very
recent ground-breaking work~\cite{Adams:2013vsa}, which demonstrated
the development of turbulence in gravitational perturbations in $3+1$
spacetime dimensions.  Thus, gravitational behavior in this regime is
effectively captured by a hydrodynamic analysis.

Drawing again on intuition from fluids, one should also expect on the
gravity side, behavior akin both to {\em turbulent} and {\em laminar}
flow. It is important to emphasize that these two phenomena can arise
on the fluid side irrespective of the velocity of the background flow
alone.  Rather, the behavior depends on the value of the {\em Reynolds
number},
\begin{equation}
  R\sim\frac{\rho v L}{\eta},
\end{equation}
where $\rho$, $v$, $L$, and $\eta$ are the characteristic energy
density, velocity fluctuation, distance scale, and shear viscosity,
describing the flow, respectively\footnote{\label{fn:R1}This definition of the
  Reynolds number is of a non-relativistic nature; for highly
  relativistic fluids, it would be desirable to have an improved
  definition. We will, however, use this definition in this
  paper, as we deal with relatively low fluid velocities.}.  
For high values of $R$, turbulence occurs, whereas for
small values the flow is laminar.  

These results and observations concerning the turbulent nature of
perturbed AAdS black branes appear to be in tension with the standard
expectation that such perturbations decay exponentially via {\em quasinormal
modes}~\cite{Berti:2009kk,Morgan:2009pn}.  Indeed, for small-amplitude
gravitational perturbations (which are dual to fluid flows with small
velocity fluctuations $v$), a linear analysis should be valid.  Due to
the symmetries of the black brane, a mode decomposition is then
possible.
Such an analysis indicates that the modes decay in time as radiation
is absorbed by the black brane (the only place where energy is lost,
as the AdS boundary acts a mirror).
Quasinormal modes of black branes in AdS can be grouped into three
``channels'' based upon their transformation properties under
rotations: the {\em sound}, {\em shear}, and {\em scalar} channels.
The longest lived families of modes within the sound and shear
channels, in turn, are known as the ``hydrodynamic'' quasinormal modes
of the black brane~\cite{Kovtun:2005ev}.  The dual fluid captures the
behavior associated with these quasinormal modes.  (Of course, the
fluid satisfies nonlinear equations, whereas quasinormal modes are
solutions to linear equations.)  Higher quasinormal modes decay more
rapidly, and are neglected in the gravity/fluid correspondence. 

Of course, any behavior dual to quasinormal mode decay is surely
absent in analyses of fluids with vanishing viscosity.  In this paper,
in order to examine this issue more closely, we extend the previous
analysis of Ref.~\cite{Carrasco:2012nf} to include viscosity, which
captures the role of the black brane as a sink of energy.  We
numerically study turbulent (and laminar) solutions for the viscous
relativistic conformal fluid (in $d=2+1$) which arises in the
gravity/fluid correspondence.  We then contrast our results with the
expectations we have laid out for the gravity dual, and we draw
conclusions about the regime of applicability of linear perturbation theory about black holes.

In the following section we review the
gravity/fluid correspondence in more detail.  We sketch the
perturbative derivation from general relativity, and we write down the
relevant equations for our work.  The dissipative relativistic
hydrodynamic equations are closely related to those of Israel and
Stewart~\cite{Israel:1976tn,Israel:1976z,Israel:1979wp}, and are thus
suitable for numerical implementation~\cite{Baier:2007ix}.  In
Sec.~\ref{sec:simulations} we proceed to describe our numerical setup,
as well as the initial data.  We work in Minkowski spacetime on
$\mathbb{R}\times T^2$, which is dual to a (periodically identified)
black brane\footnote{Similar results are expected to hold for fluids
  dual to black holes in global AdS, as already indicated in
  Ref.~\cite{Carrasco:2012nf}.} in a Poincar\'e patch of AdS.  Our
initial data consists of a shear flow, which corresponds on the
gravity side to a hydrodynamic shear quasinormal mode of the black
brane.  Due to the presence of viscosity, the shear flow is expected
to decay exponentially in the absence of turbulence, until the fluid
reaches an equilibrium state, corresponding to a (uniformly boosted)
black brane.

We present our results in Sec.~\ref{sec:results}.
Our simulations confirm that turbulent behavior and the inverse
cascade continue to manifest beyond a critical Reynolds number $R_c$,
which we determine numerically.  For $R>R_c$, the background decaying
shear flow is linearly unstable to perturbations.  Such perturbations
can grow until they reach the amplitude of the background shear flow, at
which point fully developed turbulence is attained.  
As in the inviscid case \cite{Carrasco:2012nf}, an
inverse cascade of energy is observed, eventually leaving two large
counter-rotating vortices.  On the other hand, for $R<R_c$, the shear
flow is stable to perturbations, and it decays exponentially.

Finally, in Sec.~\ref{sec:final} (as well as Appendices
\ref{app:radialmap} and \ref{app:geoquant}) we use the gravity/fluid
correspondence to relate our results for the fluid to the AAdS black
brane.  The case of laminar shear flow corresponds directly to the
ordinary decay of the hydrodynamic shear quasinormal mode of the black
brane.  However, for $R>R_c$, the instability of the fluid flow
corresponds to an instability of the quasinormal mode.  (We stress
that this does not imply an instability of the black brane, since an
overall decay continues to occur.)  Once the growing mode becomes of
order the original quasinormal mode, the original decay is interrupted
and the overall behavior is strongly modified by a fully developed
turbulent behavior.  In the 4-dimensional bulk, the energy cascades to
the longest wavelength that fits within our torus\footnote{We expect
  that for black holes, as opposed to black branes, this corresponds
  to a transfer of energy to the lowest $l$-mode. Such behavior has
  already been anticipated by the analysis
  in~\cite{Carrasco:2012nf}.}.

We conclude that ordinary perturbation theory about the uniform black
brane background is not the most suitable method of analysis for
capturing such effects analytically.  In fact, the instability is only
clearly apparent if one linearizes the Einstein equation about the
decaying quasinormal mode solution itself.  (Perturbation theory about
the uniform black brane would have to be implemented to higher orders
before the exponential growth could be recognized.)  Physically, the
reason for this behavior is that, for high-temperature black branes in
AdS, the lowest lying quasinormal modes become very long lived.  Thus,
for a given perturbation, as the temperature is increased, the linear
viscous damping term becomes small compared with nonlinear terms in
the Einstein equation.  The regime of applicability of linear
perturbation theory is thus pushed to very small metric
perturbations. On the fluid side, such properties are conveniently
captured by the Reynolds number (although, as noted above, a
relativistic generalization is desirable for relativistic fluids).
Thus it would be very interesting to obtain a geometrical realization
of the Reynolds number, in order to predict the onset of turbulence in
gravity~\cite{Evslin:2012zn}.

More generally, the unstable nature of certain long-lived quasinormal
modes suggests that the decay of a sufficiently perturbed black brane
can deviate from the picture suggested by ordinary
perturbation theory.  Rather than being describable by quasinormal
decay, the black brane can undergo a turbulent cascade with a
power law decay.  Only after the energy cascades to long wavelengths
will a quasinormal mode decay take hold.

In this work, we follow all notation and sign conventions of
\cite{Wald:1984}.  We use lower case Greek letters $(\mu,\nu =
0,1,\ldots,d-1)$ for indices of boundary quantities, and we upper case
Latin letters $(M,N = 0,1,\ldots,d)$ in the bulk.  Boundary indices
are raised and lowered with the boundary Minkowski metric
$\eta_{\mu\nu}$.

\section{Gravity/Fluid Correspondence}\label{sec:fluid}

In this section we review the basic results of the gravity/fluid
correspondence.  We sketch the derivation from Einstein's equation in
the bulk.  We also discuss issues concerning the well-posedness of
viscous relativistic fluids, and we write down suitable equations of
motion that will be used in our simulations~\cite{Baier:2007ix}.  The
derivation which we review below follows that of Bhattacharyya et al
\cite{Bhattacharyya:2008jc}.

As noted in the introduction, we restrict to boundary fluids in
Minkowski spacetime, which are dual to perturbed AAdS black branes.
Our simulations adopt $d=3$, but in this section we keep $d$
arbitrary.  We also take the boundary manifold to be $\mathbb{R}\times
T^2$; that is, we impose periodic boundary conditions along boundary
spatial directions.  Results in $d=3$ were derived in
Ref.~\cite{VanRaamsdonk:2008fp}, while the arbitrary $d$ case whose
equations we write down was analyzed in
Refs.~\cite{Haack:2008cp,Bhattacharyya:2008mz}.

The starting point for the derivation of the gravity/fluid
correspondence is a uniform boosted black brane spacetime, which, written in
ingoing Eddington-Finkelstein coordinates, reads,
\begin{equation}\label{eq:metric0}
  ds^2_{[0]} = -2 u_\mu dx^\mu dr + 
 r^2 \left( \eta_{\mu\nu} + \frac{1}{(b r)^d} u_\mu u_\nu  \right) dx^\mu dx^\nu.
\end{equation}
Here, the fields $b$ and $u_\mu$ (satisfying $u^\mu u_\mu = -1$) are constants. This is a solution to the bulk Einstein equation,
\begin{equation}
  G_{AB} + \Lambda g_{AB} = 0,
\end{equation}
with the cosmological constant $\Lambda = - d(d-1)/2$.  The boosted
black brane is related to the static black brane by a coordinate
transformation.  The coordinates $x^\mu = (t,x,y)$ are to be thought
of as ``boundary'' coordinates, while the coordinate $r$ is the
``bulk'' radial coordinate.

To each asymptotically AdS bulk solution, is associated a metric and
conserved stress-energy tensor on the timelike boundary of the
spacetime at $r\to\infty$ (see, e.g.,
Ref.~\cite{Balasubramanian:1999re} or Appendix~\ref{app:radialmap} for
the precise definition).  The boundary metric, in the case of
\eqref{eq:metric0} is $\eta_{\mu\nu}$, while the boundary
stress-energy tensor is
\begin{equation}
  T_{\mu\nu}^{[0]} = \frac{1}{16\pi G_{d+1} b^d}(du_\mu u_\nu + \eta_{\mu\nu}).
\end{equation}  
This is a fluid stress-energy tensor, so one may read off the energy
density and pressure,
\begin{align}
  \rho &= \frac{d-1}{16\pi G_{d+1} b^d},\\
  P &= \frac{1}{16\pi G_{d+1} b^d}.
\end{align}
The stress-energy tensor is traceless, with equation of state
\begin{equation}
  P=\frac{\rho}{d-1},
\end{equation}
as required by conformal invariance.  Imposing the first law of
thermodynamics, $\mathrm{d}\rho = T \mathrm{d}s$, as well as the
relation $\rho+P = sT$, gives the entropy density $s$
and fluid temperature $T$,
\begin{align}
  s&=AT^{d-1},\\
  \rho&=\frac{d-1}{d}AT^d.
\end{align}
Here, $A$ is a constant of integration.  This is fixed to $A \equiv
(4\pi)^d/(16\pi G_{d+1} d^{d-1})$ by equating $T$ with the Hawking
temperature $T_{Hawking} = d / (4\pi b)$ of the black
brane\footnote{Our simulations use units where $A=1$.}.

To move beyond the uniform fluid, $b$ and $u_\mu$ are promoted to
functions of the boundary coordinates $x^\mu$, which are slowly
varying; that is, if $L$ is the length scale of variation of these
fields, then $L\gg b$.  At this point, the metric \eqref{eq:metric0}
is no longer a solution to Einstein's equation.  However, due to the
fact that the fields are {\em slowly} varying, it is possible to
systematically correct the metric order by order in a derivative
expansion, so that Einstein's equation is solved to any given order in
derivatives.  One can then compute the boundary stress-energy tensor
corresponding to the metric at each order, and take this as defining
the boundary fluid.

In this setup, the boundary metric is fixed to $\eta_{\mu\nu}$
throughout.  This can be thought of as a ``Dirichlet condition'' on
the boundary.  The Einstein equation reduces to a set of
``constraints'' along the timelike boundary of AdS, as well as
evolution equations into the bulk.  The ``momentum constraint'' gives
rise to conservation of boundary stress-energy, while the
``Hamiltonian constraint'' ensures tracelessness.  The ``evolution
equations'' reduce to ordinary differential equations along $r$.
Regularity at the future black brane horizon is imposed as one of the
boundary conditions for these ODEs.  This corresponds to the
imposition of an ingoing boundary condition, and it is responsible for
the breaking of time-reversal symmetry inherent in the fluid viscosity
term.  The ``Landau frame'' gauge condition,
\begin{equation}
  u^\nu T_{\mu\nu} \propto u_\mu,
\end{equation}
is also imposed.

After a rather long, but direct, calculation, the resulting
boundary stress-energy tensor -- to second order in derivatives -- is
found to be
\begin{equation}\label{eq:Tmunu2}
  T_{\mu\nu}^{[0+1+2]} = \frac{\rho}{d-1}\left(d u_\mu u_\nu + \eta_{\mu\nu}\right) + \Pi_{\mu\nu},
\end{equation}
where the viscous part, $\Pi_{\mu\nu}$, is (see Eq.~(3.11) of
Ref.~\cite{Baier:2007ix}, Eq.~(1.5) of Ref.~\cite{Haack:2008cp}, or
Eq.~(1.3) of Ref.~\cite{Bhattacharyya:2008mz})
\begin{align}\label{eq:Pi1}
  \Pi_{\mu\nu}={}&- 2\eta \sigma_{\mu\nu} \nonumber\\
  &+ 2\eta\tau_\Pi\left(\langle u^\alpha\partial_\alpha \sigma_{\mu\nu}\rangle + \frac{1}{d-1}\sigma_{\mu\nu}\partial_\alpha u^\alpha\right)+ \langle \lambda_1 \sigma_{\mu\alpha}\sigma_{\nu}^{\phantom{\nu}\alpha} + \lambda_2 \sigma_{\mu\alpha}\omega_{\nu}^{\phantom{\nu}\alpha} + \lambda_3 \omega_{\mu\alpha}\omega_{\nu}^{\phantom{\nu}\alpha}\rangle.
\end{align}
The shear and vorticity tensors are given by
\begin{align}
  \sigma_{\mu\nu} &\equiv \langle\partial_\mu u_\nu\rangle,\\
  \omega_{\mu\nu} &\equiv P_\mu^{\phantom{\mu}\alpha}P_\nu^{\phantom{\nu}\beta}\partial_{[\alpha}u_{\beta]}.
\end{align}
The angled brackets denote the symmetric traceless part of the projection
orthogonal to $u^\mu$,
\begin{equation}
  \langle A_{\mu\nu}\rangle \equiv \left(P_{(\mu}^{\phantom{(\mu}\alpha}P_{\nu)}^{\phantom{\nu)}\beta} - \frac{1}{d-1}P_{\mu\nu}P^{\alpha\beta}\right)A_{\alpha\beta},
\end{equation}
while $P_{\mu\nu}$ is the spatial projector orthogonal to $u^\mu$,
\begin{equation}
  P_{\mu\nu} \equiv \eta_{\mu\nu} + u_\mu u_\nu.
\end{equation}
It may be verified that $\Pi_{\mu\nu}$ is symmetric and satisfies
\begin{align}
  \Pi^\mu_{\phantom{\mu}\mu}&=0,\\
  u^\nu\Pi_{\mu\nu}&=0.
\end{align}
The transport coefficients $\{\eta,\,\tau_\Pi,\,\lambda_i\}$ have been
worked out explicitly in various dimensions
\cite{VanRaamsdonk:2008fp,Haack:2008cp,Bhattacharyya:2008mz},
\begin{align}
  \label{eq:eta}  \eta&=\frac{s}{4\pi},\\
  \tau_\Pi&=b\left[1-\int_1^\infty\frac{y^{d-2}-1}{y(y^d-1)}\,\mathrm{d}y\right]\xrightarrow{d=3}b\left(1-\frac{1}{2}\log 3  + \frac{\pi}{6\sqrt{3}}\right),\\
  \lambda_1&=\frac{\eta b}{2},\\
  \lambda_2&=-2\eta b \int_1^\infty\frac{y^{d-2}-1}{y(y^d-1)}\,\mathrm{d}y \xrightarrow{d=3} - \eta b \left( \log 3 - \frac{\pi}{3\sqrt{3}}  \right),\\
  \label{eq:lambda3} \lambda_3&=0.
\end{align}
Thus, the boundary fluid has a nonzero shear viscosity $\eta$, but the
bulk viscosity vanishes, so that the stress-energy tensor remains
traceless.  Conformal invariance can also be used to deduce the
presence of the particular nonzero second order transport coefficients
directly~\cite{Baier:2007ix}.  For completeness, we note that
conservation of the boundary stress-energy tensor leads to the
equations of motion for $\rho$ and $u^\mu$,
\begin{align}
  \label{eq:drho}0 &= u^\mu\partial_\mu \rho + \frac{d}{d-1}\rho \partial_\mu u^\mu - u^\mu\partial^\nu \Pi_{\mu\nu},\\
  \label{eq:du}0 &= \frac{d}{d-1}\rho u^\mu\partial_\mu u^\alpha +
  \frac{1}{d-1} \partial^\alpha\rho -
  \frac{d}{(d-1)^2}u^\alpha\rho\partial_\mu u^\mu +
  \frac{1}{d-1}u^\alpha u^\mu \partial^\nu \Pi_{\mu\nu} +
  P^{\alpha\mu}\partial^\nu\Pi_{\mu\nu}.
\end{align}

It is easy to see that in this derivative expansion, $\Pi_{\mu\nu}$ is
subleading in $b/L$, as compared with the perfect fluid stress-energy
tensor $T_{\mu\nu}^{(0)}$.  Thus, given a fixed $L$, the viscous part
may be neglected for small $b$, or, equivalently, large $T$.  This is
the limit which was taken in Ref.~\cite{Carrasco:2012nf}. In this
work, however, we wish to move beyond the $T\to\infty$ limit, so
viscosity must be included in our simulations.  As explained in the
introduction, this corresponds on the gravity side to the effects of
energy losses through the horizon.

At this point, one may wonder why we have bothered to include terms to
{\em second order} in derivatives, since the shear viscosity appears
at first order.  The reason is that relativistic viscous fluid
formulations which are first order in derivatives, as originally laid
out by Eckart~\cite{Eckart:1940}, lead to acausal propagation, and are
generally ill-posed~\cite{Israel:1976tn}.  It turns out to be possible
to resolve these issues and to produce a hyperbolic system by
including second order terms; in particular, the term involving
$\tau_\Pi$~\cite{Israel:1976z,Israel:1976tn,Israel:1979wp,Hiscock1983466}. The
second order terms which appear in Eq.~\eqref{eq:Pi1} resolve the
problems introduced by the viscosity, but they bring about analogous
issues at higher order. To fully resolve these difficulties, it is
necessary to promote $\Pi_{\mu\nu}$ to an independent field.  Then, one
reduces the order of the system of equations by substituting $ - 2 \eta
\sigma_{\mu\nu} \to \Pi_{\mu\nu}$ on the second line of
\eqref{eq:Pi1}.  This substitution is consistent to the order to which
we are working in the derivative expansion.  Furthermore, this
assumption will remain valid in $2+1$ dimensions under time evolution
by virtue of the expected inverse
cascade~\cite{Carrasco:2012nf}. Therefore, following Baier et
al~\cite{Baier:2007ix}, we obtain
\begin{align}\label{eq:Pi}
  \Pi_{\mu\nu} ={}& -2\eta \sigma_{\mu\nu}\nonumber\\
  & - \tau_\Pi \left(\langle u^\alpha \partial_\alpha \Pi_{\mu\nu}\rangle + \frac{d}{d-1}\Pi_{\mu\nu} \partial_\alpha u^\alpha\right) + \langle \frac{\lambda_1}{\eta^2}\Pi_{\mu\alpha}\Pi_\nu^{\phantom{\nu}\alpha} - \frac{\lambda_2}{\eta} \Pi_{\mu\alpha}\omega_\nu^{\phantom{\nu}\alpha} + \lambda_3 \omega_{\mu\alpha}\omega_\nu^{\phantom{\nu}\alpha}\rangle.
\end{align}

The formulation we have described above also includes additional
second order terms with coefficients $\{\lambda_i\}$.  We have decided
to include these in the interest of completeness, although we find
that they have no effect on our results.  Indeed, as discussed by
Geroch \cite{Geroch:1995bx,Geroch:2001xs}, all hyperbolic relativistic
theories of fluids with viscosity should be {\em physically
  equivalent}.  By this, one means that any additional terms in the
equations of motion, when evaluated within the domain of applicability
of the theory, should be small, as compared with the lower order
terms.  That is, if the higher order terms became important, then
there would be no justification in not including even higher order
terms, and the perturbative expansion would break down.  This also
means that the specific value of $\tau_\Pi$ is unimportant, so long as
it is sufficiently large that the theory is causal.  (We will use this
fact later to increase its value in order to speed up our numerical
simulations.)

To summarize, the system of interest is described by
Eqs.~\eqref{eq:drho}, \eqref{eq:du}, and \eqref{eq:Pi}. These
equations, however, require further manipulation prior to numerical
implementation.  For comparison, we recall that in the context of
inviscid hydrodynamics, it is covenient to
express the hydrodynamic equations in {\em conservation form} (see,
e.g., \cite{lev02}).  That is, $t$-derivatives of energy and momentum
density are equated with $x^i$-derivatives of fluxes.  Such a form of
the equations is particularly advantageous when studying solutions that
can develop sharp gradients or discontinuities, and was employed in
Ref.~\cite{Carrasco:2012nf}.  However, a simple extension of this
approach is not possible in the presence of viscosity since the
evolution equation \eqref{eq:Pi} for $\Pi_{\mu\nu}$ is not of the
desired form.  In particular, since $P_{\mu\nu}$ projects orthogonally
to $u^\mu$ rather than to $\partial_t^\mu$, this equation
contains a $t$-derivative of $u^\mu$ in addition to that of
$\Pi_{\mu\nu}$. However, since the presence of viscocity prevents the
development of steep gradients and discontinuities in our solutions,
adopting a conservative form is not necessary.

We therefore follow an approach similar to that employed
in~\cite{Luzum:2008cw,Romatschke:2009im} within the context of heavy
ion collisions.  To begin, we note that in $d=3$ the conditions $u_\mu
u^\mu = -1$, $u^\nu\Pi_{\mu\nu} = 0$, and $\Pi^\mu_{\phantom{\mu}\mu}
= 0$ reduce the number of dynamical variables to five.  We take these
to be 
$\mathcal{U}\equiv (\rho,u_x,u_y,\Pi_{xx},\Pi_{xy})$.
Equations~\eqref{eq:drho}, \eqref{eq:du}, and \eqref{eq:Pi} are quite
complicated when expressed in terms of $\mathcal{U}$, and in order to
evolve the equations numerically, one must solve for the
$t$-derivatives of the fields.
Using computational algebra software, we can write our equations in the desired form,
\begin{equation}\label{eq:numeq}
  \partial_t\mathcal{U} = \mathcal{F}(\mathcal{U},\partial_i \mathcal{U});
\end{equation}
and these are the equations we implement in our code.

\section{Simulations}\label{sec:simulations}

In this section we describe our choice of initial data and details of
our numerical setup.

\subsection{Initial data}\label{sec:initialdata}

As described in the introduction, we choose initial data corresponding
to a shear hydrodynamic quasinormal mode of the black brane. Our studies
concentrate on nonlinear phenomena described by the system; however, for 
future reference, in this
subsection we analyse the evolution under the linearized equations of motion.  

Consider perturbations about a uniform fluid solution,
\begin{align}
  \rho_{(0)} &= \text{constant},\\
  u_{(0)}^\mu &= (1,0,0),\\
  \Pi^{(0)}_{\mu\nu} &= 0,
\end{align}
which is dual to a non-boosted uniform black brane in the bulk.
Solutions to the {\em linearized} equations of motion whose only
nonzero perturbed fields are $u_{(1)}^x = u_{(1)}^x(t,y)$ and
$\Pi^{(1)}_{xy} = \Pi^{(1)}_{yx} = \Pi^{(1)}_{xy}(t,y)$ describe {\em
  shear flow}.  That is, fluid flow orthogonal to the velocity
gradient \cite{Baier:2007ix}.  The linearized equations of motion for
shear flow reduce to
\begin{align}
  \label{eq:linux}0 &= \frac{3}{2} \rho_{(0)} \partial_t u^x_{(1)} + \partial_y\Pi^{(1)}_{xy},\\
  \label{eq:linPi}\Pi_{xy} &= -\eta_{(0)} \partial_y u_x^{(1)} - \tau_\Pi^{(0)}\partial_t \Pi^{(1)}_{xy},
\end{align}
which may be solved by expanding in Fourier modes.
Considering one mode, with spacetime dependence the form $\sim
e^{-i\omega t + i k y}$, Eqs.~\eqref{eq:linux} and \eqref{eq:linPi}
give \cite{Baier:2007ix}
\begin{equation}
  0 = \omega^2 + \frac{i}{\tau_\Pi^{(0)}} \omega - \frac{2k^2\eta_{(0)}}{3\rho_{(0)}\tau_\Pi^{(0)}}.
\end{equation}
This has two solutions for small $k$,
\begin{equation}\label{eq:shearfrequencies}
  \omega_1 \approx - i \frac{2k^2\eta_{(0)}}{3\rho_{(0)}} = -i\frac{k^2}{4\pi T_{(0)}},\qquad \omega_2\approx -\frac{i}{\tau_\Pi^{(0)}},
\end{equation}
both of which describe pure exponential decay.  The first solution
corresponds, in the bulk, to the hydrodynamic shear quasinormal mode of the
black brane \cite{Baier:2007ix,Bhattacharyya:2008jc,Kovtun:2005ev}.
(The second solution shows that $\tau_\Pi$ is the decay timescale for
$\Pi_{\mu\nu}$ to approach $-2\eta\sigma_{\mu\nu}$.)

Thus, since we are interested in understanding the corresponding black
brane quasinormal mode in a nonlinear context, we choose initial data
with
\begin{align}
  \rho(t=0) &= \rho_0 = \text{constant}, \\
  u_x(t=0) &= v_0 \sin\left(\frac{2\pi n y}{D}\right),
\end{align}
and all other fields zero\footnote{This initial data was also chosen
  in the Appendix of Ref.~\cite{Carrasco:2012nf} because, in the
  inviscid case, it leads to a stationary solution.}. We vary the
background energy density $\rho_{0}$, velocity amplitude $v_0$, the
number of modes $n$, and the torus size $[0,D]^2$.  The reason $n$ and
$D$ are varied separately (rather than as the wavelength $\lambda =
D/n$) is that effects due to the finite size of the box can come into
play for small $n$.

Moving from the linear to nonlinear level, we expect the pure decay of
this shear flow to persist, at least for small velocities.  That this
is the case will be verified in Sec.~\ref{sec:results}.  We also keep
the velocities small in most of our simulations in order to match to
the linear predictions.

Our main reason for setting up this flow, of course, is to study its
stability.  In order to do so,
we also initially seed $u_x$ with a very small random
perturbation\footnote{In Ref.~\cite{Carrasco:2012nf}, this same effect
  was achieved via small numerical errors.}.  By studying the effects
of this perturbation, one can learn about the robustness (or lack
thereof) of pure quasinormal mode decay.

\subsection{Numerical setup}

Equation \eqref{eq:numeq} was solved numerically using the method of
lines.  To perform the spatial discretization, fourth order accurate
spatial derivatives were used, while third order Runge-Kutta was used
for time integration (see,
e.g.,~\cite{Calabrese:2003yd,Calabrese:2003vx}).  Consequently, third
order convergence is expected when turbulence does not arise. To
confirm that this is the case, we studied laminar flows (with
$\rho_0=10^7$, $v_0=0.02$, $n=10$ and $D=10$), and adopted grid
spacings $\Delta x_N = \Delta y_N = 0.1/N$ with $N=1,2,4$.  We
computed the convergence rate
$\|\mathcal{U}(N=1)-\mathcal{U}(N=2)\|_2/\|\mathcal{U}(N=4)-\mathcal{U}(N=2)\|_2\equiv
2^{p}$, and found $p\approx3$.  Typical simulations were thus performed on
$201\times201$ grids (with periodic boundary conditions).  With the
torus size $D$ (typically, $D=10$), the corresponding grid spacing
was then $\Delta x = \Delta y = (D/10)\times0.05$.

We note that the presence of the short viscous timescale $\tau_\Pi$
[see Eq.~\eqref{eq:shearfrequencies} in the previous subsection]
imposes a harsh constraint on the timestep length for an explicit
integration method,
\begin{equation}
  \Delta t  \propto \tau_\Pi.
\end{equation}
For our simulations, this is a much stronger constraint than that
arising from the finite propagation speeds of the solution (i.e., the
CFL condition).  However, as we discussed in Sec.~\ref{sec:fluid}, the
precise value of $\tau_\Pi$ should not have physical significance, as
long as the equations of motion remain hyperbolic.  Therefore, to
allow for a more efficient numerical integration, we increased
$\tau_\Pi$ by a factor of 100 for many of our runs.  We verified that
this had no significant effects on any of the physical properties we
measured.

\section{Results and Analysis}\label{sec:results}

In this section we present and analyze results.  We first define the
Reynolds number for the shear flow, which we find to accurately
predict the onset of instability.  We then describe the three observed
``phases'' of a fully developed turbulent flow: initial growth of
instabilities, inverse turbulent energy cascade, and final exponential
decay (see Fig.~\ref{fig:turbulence-xsection} for a preview).  Our
results are largely consistent with expectations drawn from solutions
to the Navier-Stokes equations in $2+1$ dimensions.

\subsection{Reynolds number}

For steady flows, the Reynolds number is a useful dimensionless
quantity which can be used to predict stability (see, e.g.,~\cite{Landau:1987}).
It is generally true that for sufficiently low Reynolds numbers, the
flow is stable with respect to small perturbations, in which case it
is said to be {\em laminar}.  In contrast, for large Reynolds numbers
the flow is unstable, which eventually leads to {\em turbulence}.  The
{\em critical} Reynolds number which separates these two regimes
depends upon the particular flow under consideration.

The shear flow that we study is {\em not} steady, causing the Reynolds
number to change with time.  [The shear viscosity causes it to decay
according to Eq.~\eqref{eq:shearfrequencies}.]  Nevertheless, it is 
a useful quantity to
consider, as the decay can be treated in a quasi-stationary
manner\footnote{For non-steady flows, one can also consider the {\em
    Strouhal number}, but this adds nothing new unless the fluid is
  externally forced \cite{Landau:1987}.}.  In this case, the stability
properties of the flow depend only upon the instantaneous value of the
Reynolds number.

For our flow, we define the Reynolds number to be\footnote{As noted in
  Footnote \ref{fn:R1}, it would be desirable to define a Reynolds number
  suitable for highly relativistic flows.  However, we restrict to
  small velocities, so Eq.~\eqref{eq:Rnumber} is adequate for our
  purposes.}
\begin{equation}\label{eq:Rnumber}
 R \equiv \frac{\rho \lambda}{\eta}\max(u_x).
\end{equation}
Substituting for $\eta$ and $\lambda$,
\begin{equation}
R = 6\pi T \frac{D}{n}\max(u_x) = \left(\frac{3}{2A}\right)^{1/3}6\pi  \rho^{1/3}\frac{D}{n}  \max(u_x).
\end{equation}
Thus initially,
\begin{equation}
  R(t=0) = \left(\frac{3}{2A}\right)^{1/3}6\pi  \rho_0^{1/3}\frac{D}{n} v_0,
\end{equation}
and with time, $R$ decays with $\max(u_x)$.  ($\rho$, $D$ and $n$ are
all either constants, or nearly constant.)  Later in this section, we
will verify that there exists a critical Reynolds number $R_c$,
and we will determine its value.

Strictly speaking, flows which have different values of $n$ are {\em
  not} geometrically similar, meaning that they are not related by a
universal scaling of distances.  This is due to the presence of two
independent associated length scales (the wavelength $\lambda$, and
box size $D$).  So, one should exercise caution when comparing 
the Reynolds numbers of two such flows.  However, for large values of
$n$, the finite box size should not play an important role in
governing stability, and our definition \eqref{eq:Rnumber} makes
sense.  (We will address effects at small $n$ later in this section.)
Our definition of $R$, and our decision to compare flows at different
$n$, is motivated both by simplicity, as well as the desire to address
the infinite brane limit ($n,D\to\infty$ while holding $\lambda$
fixed).

We note that, for fluids dual to black holes with compact spatial
sections, one should also be careful when using a value of $R_c$ for
high angular quantum number flows, to predict stability of low-$l$
flows.  In particular, as we will discuss further, we would expect
$l=2$ shear modes to be stable for any value of $R$.

\subsection{Stability of shear flow}

In this subsection we analyze the early stages of the flow, where the
overall properties are governed by the shear decay in $u_x$.

Recall that, in addition to the shearing configuration, the initial
data is seeded with a small-amplitude random perturbation.  This has
the potential to grow or decay, depending on the Reynolds number of
the flow.  To track the presence of growing instabilities, we
monitored the $u_y$ field.  In the linear analysis of
Sec.~\ref{sec:initialdata}, $u_y$ remains exactly zero, so its growth
reflects the growing unstable mode.  (In addition, even for stable
flows, $u_y$ becomes nonzero due to nonlinearities; but this remains
small.)  As expected, as we
varied the initial data we found that the various solutions could be
categorized into several groups, based on the growth of $\|u_y\|_2$.


\begin{figure}
  \centering
  \begin{tabular}{c c}
    \subfloat[$t=0$]{\label{fig:turb-a}\includegraphics[width=.48\textwidth]{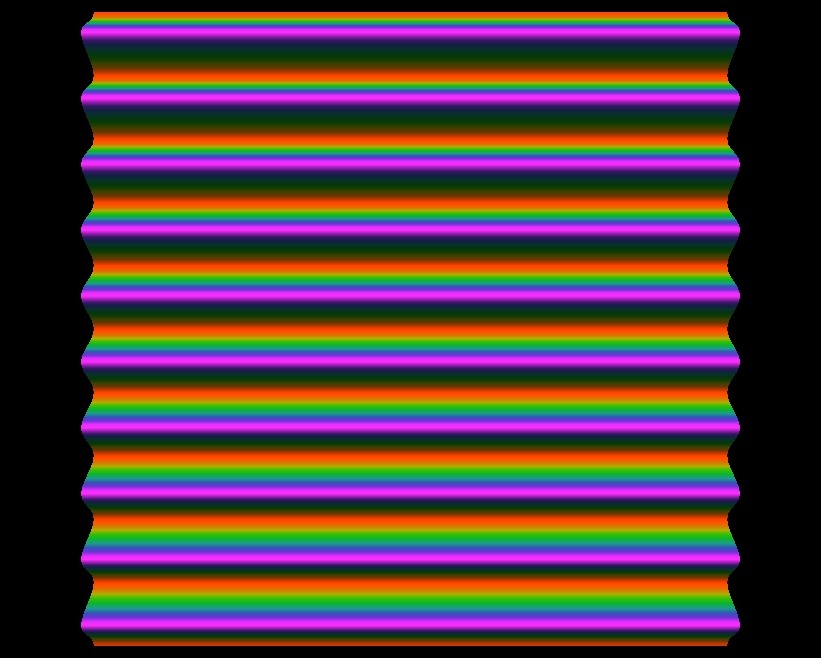}} &
    \subfloat[$t=350$]{\label{fig:turb-b}\includegraphics[width=.48\textwidth]{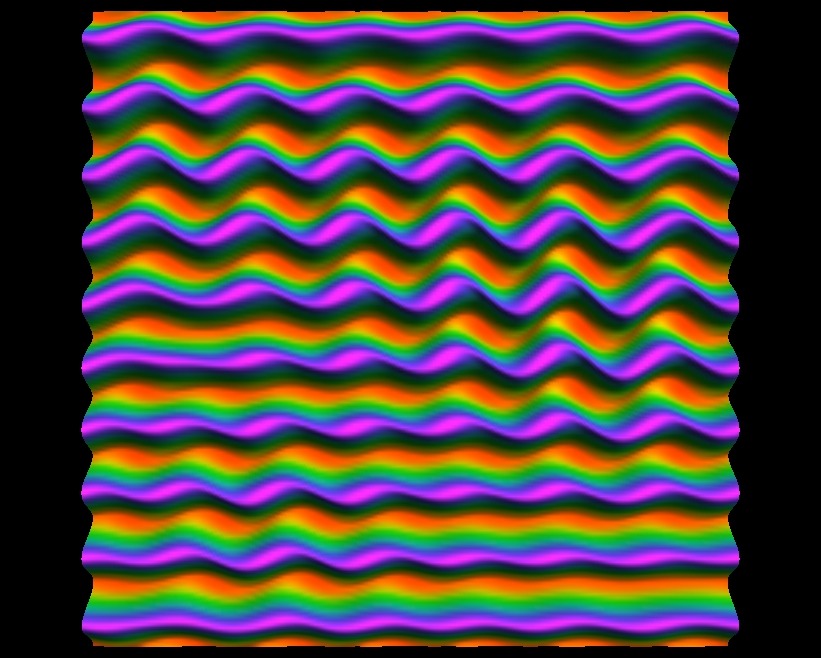}} \\
    \subfloat[$t=500$]{\label{fig:turb-c}\includegraphics[width=.48\textwidth]{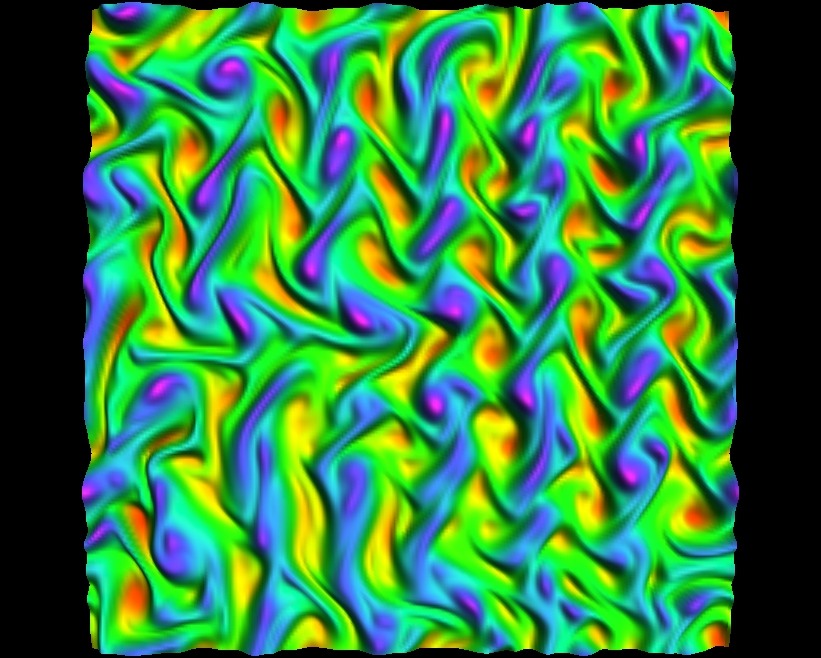}} &
    \subfloat[$t=900$]{\label{fig:turb-d}\includegraphics[width=.48\textwidth]{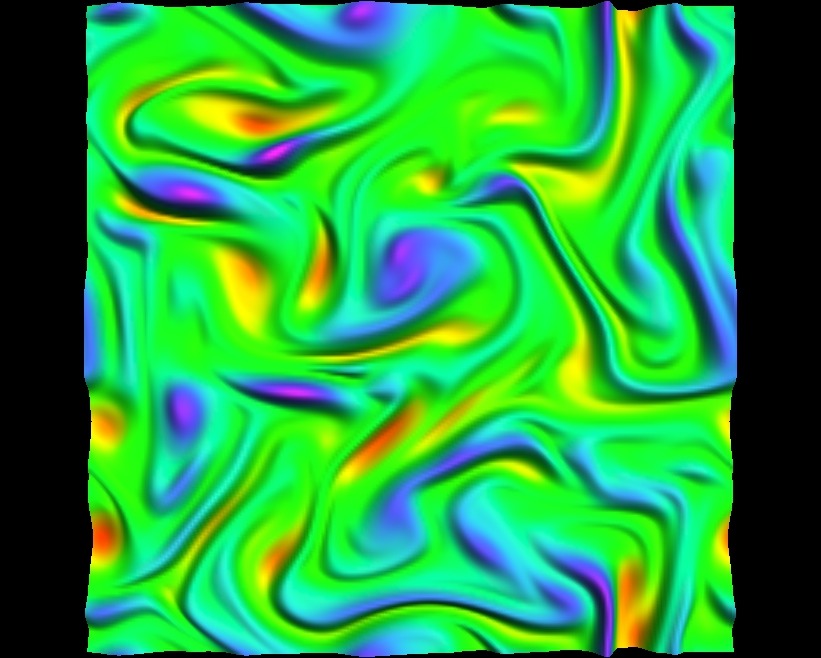}} \\
    \subfloat[$t=2500$]{\label{fig:turb-e}\includegraphics[width=.48\textwidth]{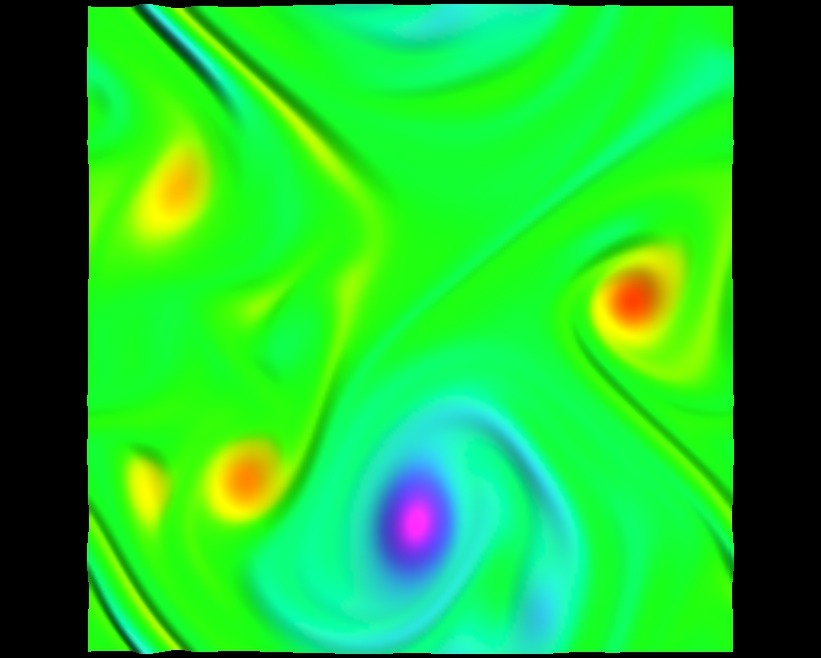}} &
    \subfloat[$t=7000$]{\label{fig:turb-f}\includegraphics[width=.48\textwidth]{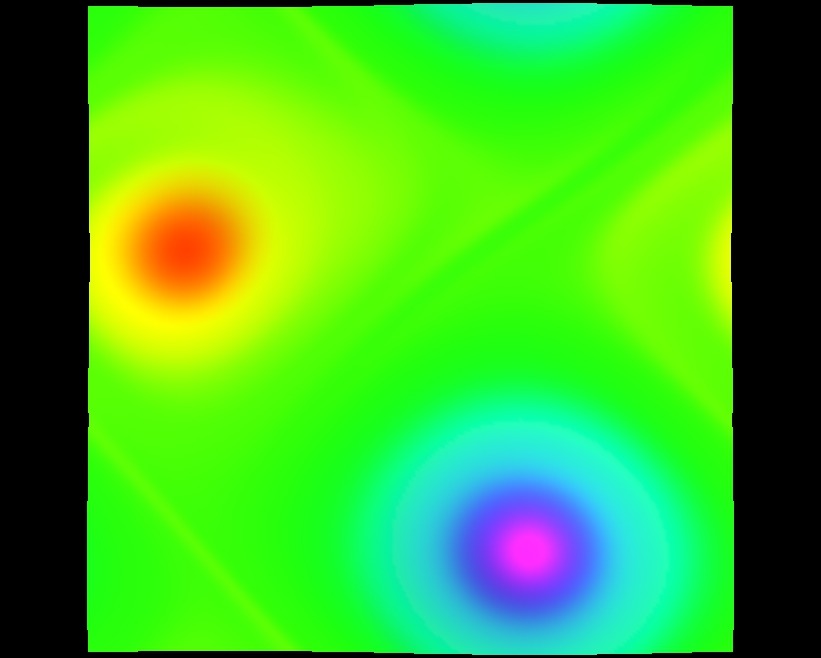}}
  \end{tabular}
  \caption{Vorticity field at various times for a turbulent run
    ($\rho_0=10^{10}$, $v_0=.05$, $D=10$, $n=10$).  The inverse
    cascade behavior is evident, leading to two counter-rotating, and
    slowly decaying, vortices at late times.}
  \label{fig:turbulence-xsection}
\end{figure}
\begin{figure}
  \centering
  \begin{tabular}{c c}
    \subfloat[Vorticity]{\label{fig:turbvort}\includegraphics[width=.5\textwidth]{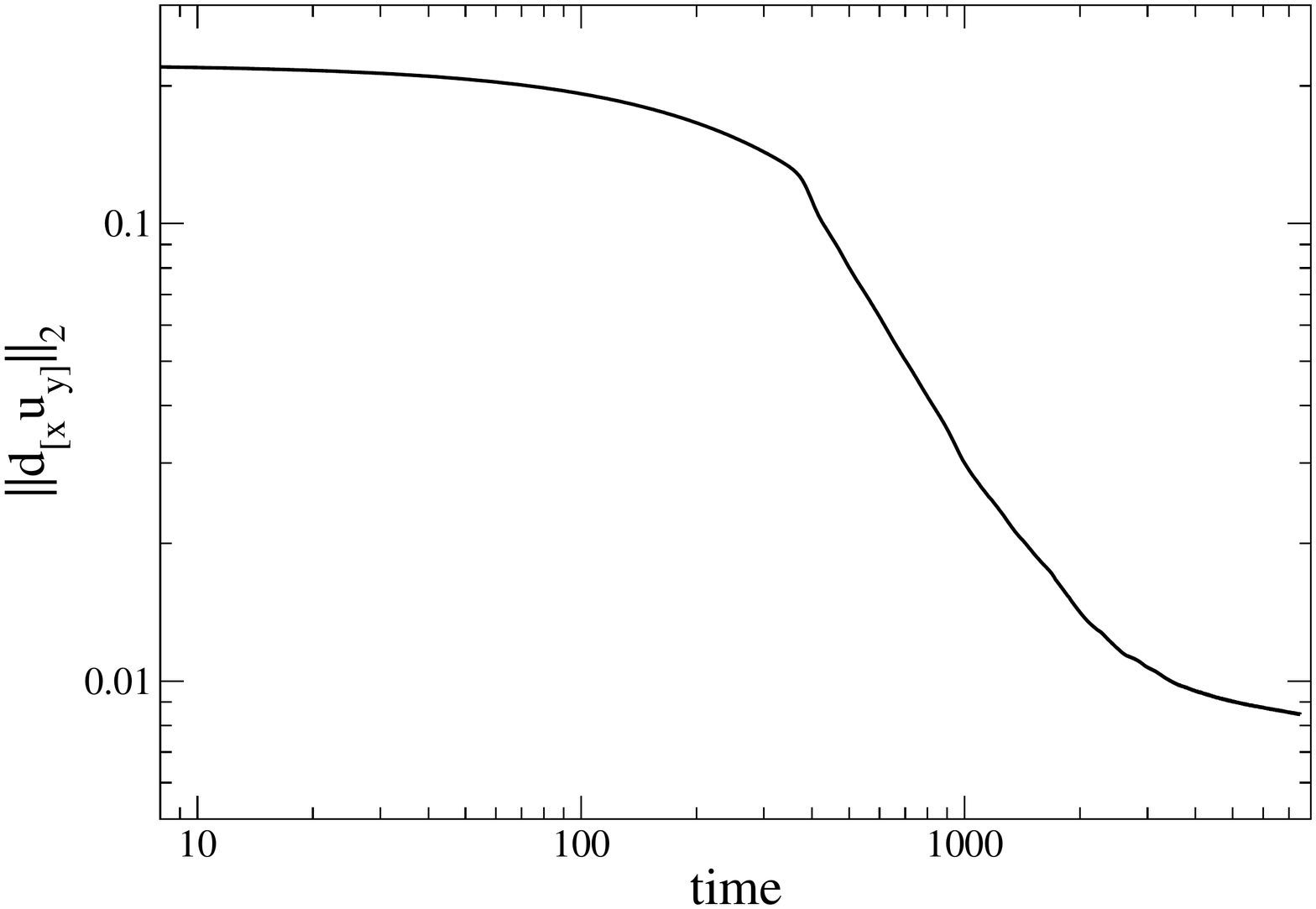}}&
    \subfloat[Velocity]{\label{fig:turbu}\includegraphics[width=.5\textwidth]{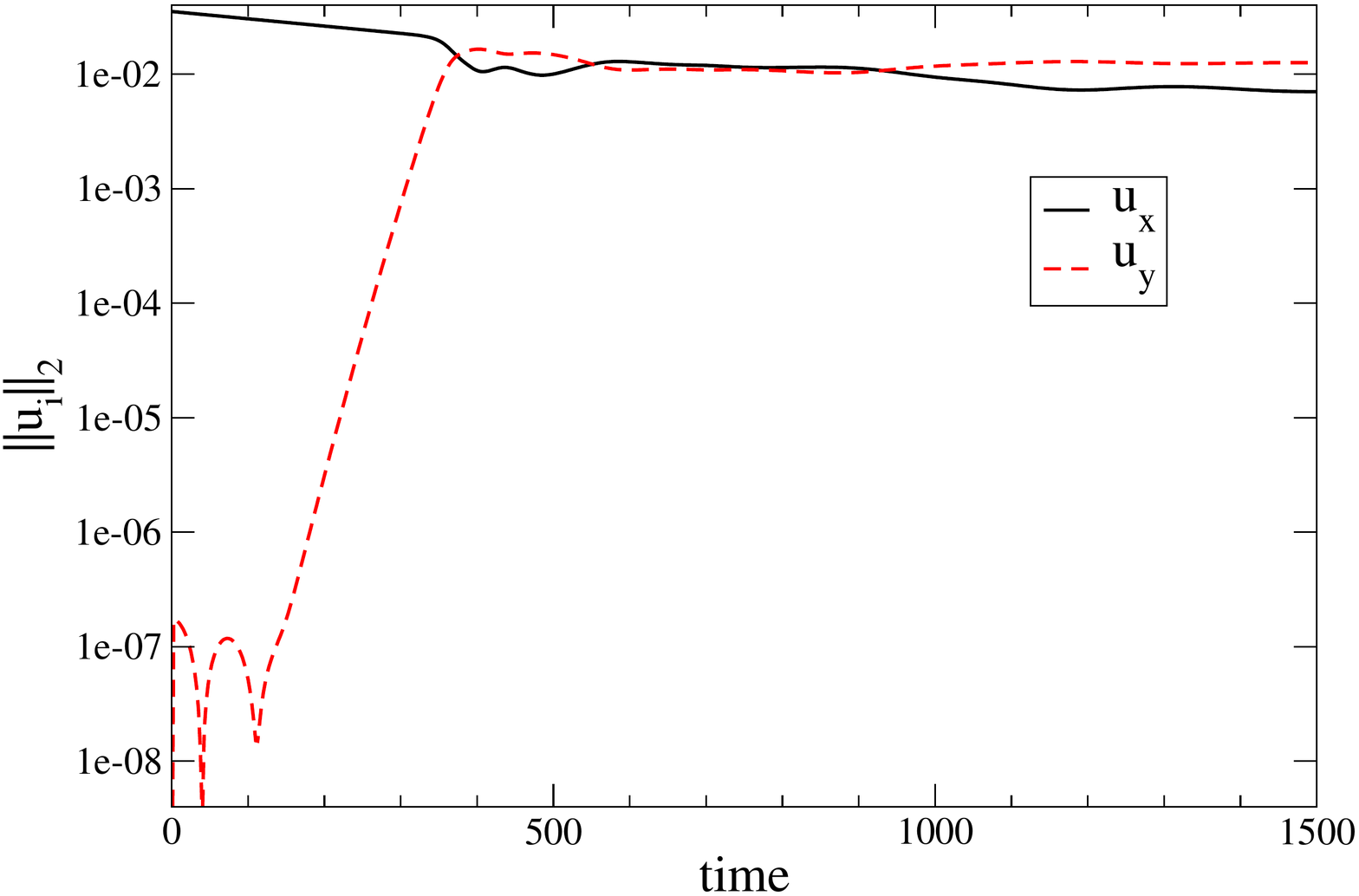}}
  \end{tabular}
  \caption{$L_2$ norms of (a) vorticity, and (b) $u_i$, as a function
    of time for turbulent flow (same run as
    Fig.~\ref{fig:turbulence-xsection}).  In (a) there is an initial
    exponential decay, followed by a power law during the inverse
    cascade, followed by a slower exponential decay.  In (b) $u_y$
    grows exponentially until it is of similar amplitude to $u_x$.}
  \label{fig:turbulence}
\end{figure}
\begin{figure}
  \centering
  \includegraphics[width=.48\textwidth]{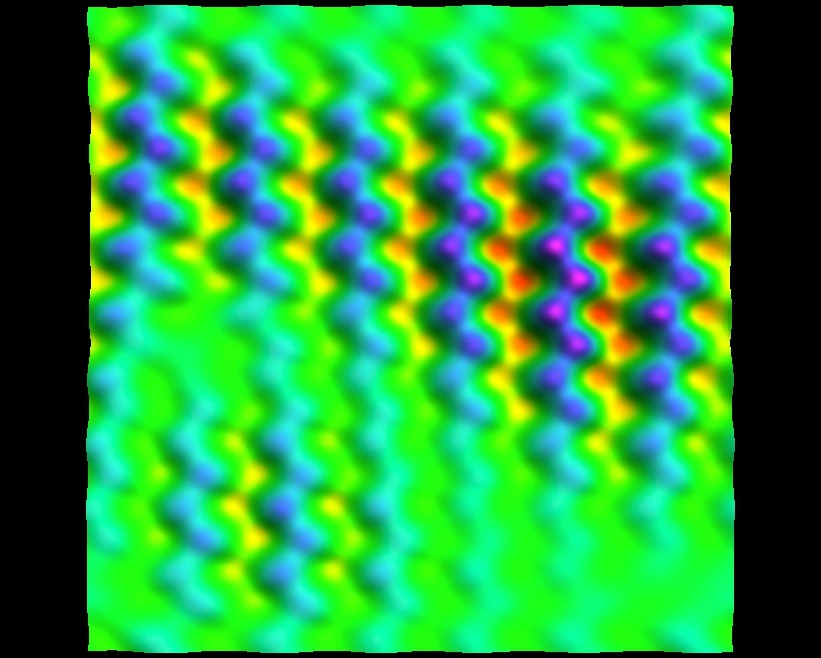}
  \caption{$u_y$ field at $t=300$ for the same run as
    Fig.~\ref{fig:turbulence-xsection}.}
  \label{fig:turbulence-uy-xsection}
\end{figure}
At one extreme, the typical turbulent run is illustrated in
Fig.~\ref{fig:turbulence-xsection}.  This shows the vorticity field at
several times.  In Fig.~\ref{fig:turbulence}, we show the
corresponding $L_2$ norms of the vorticity and the components of
$u_i$.  We see that initially, the vorticity and $u_x$ decay
exponentially in the manner expected for shear flow.  However, during
that time, $\|u_y\|_2$ undergoes exponential {\em growth}, until it
reaches the same amplitude as $\|u_x\|_2$.  This brings the solution
into an equipartition of energy between $u_x$ and $u_y$.  At
this point the initial decay has been disrupted, and turbulence sets
in, as we will describe in more detail in the following subsections.
The onset of this behavior is known as  the Kelvin-Helmholtz instability in fluid dynamics.

Visual inspection of the $u_y$ field (see
Fig.~\ref{fig:turbulence-uy-xsection}) indicates that the growing mode
itself is also very roughly a shear mode,
\begin{equation}\label{eq:growinguy}
  u_y \approx f(t)\sin\left(\frac{2 \pi m y}{D}+ \phi\right), 
\end{equation}
typically with $n/2 < m < n$.  As we have alluded to earlier, the
finite box size plays a role at small $n$, and it comes into play
here.  For example, we find that the $n=1$ case is stable, even for
large $R$, because the box does not admit a mode with $m<1$.  This is
the expected behavior as there is no room for an inverse cascade,
given an $n=1$ initial configuration. For large $n$, there is no
obstacle in fitting the growing mode into the box, and the box size
$D$ plays no role.  Thus, fixing $\lambda$ and extrapolating to the
infinite box ($D\to\infty$), the instability should be present for
sufficiently large $R$.

\begin{figure}
\centering
  \begin{tabular}{c c}
    \subfloat[Vorticity]{\label{fig:lamvort}\includegraphics[width=.5\textwidth]{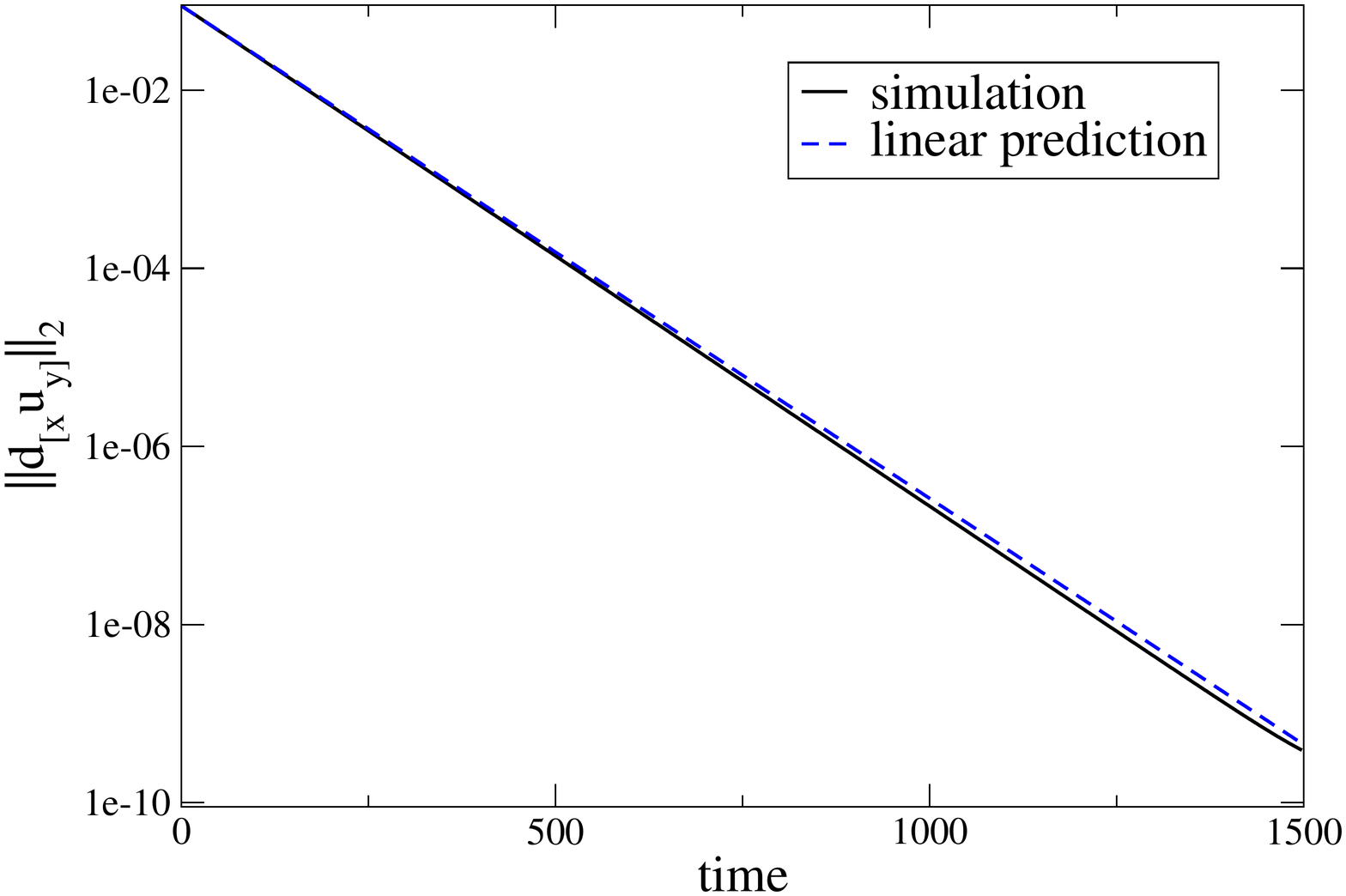}}&
    \subfloat[Velocity]{\label{fig:lamu}\includegraphics[width=.5\textwidth]{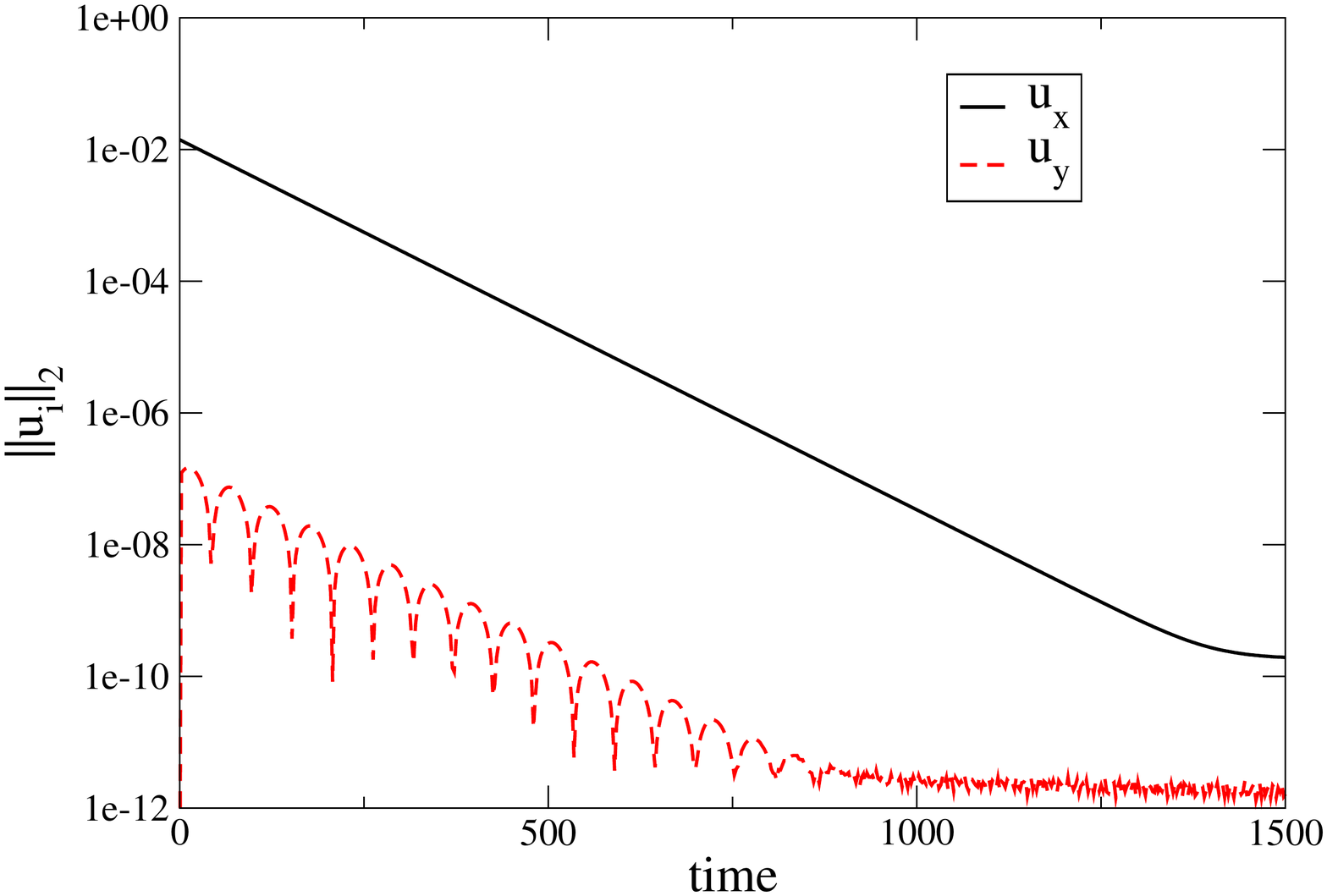}}
  \end{tabular}
  \caption{$L_2$ norms of (a) vorticity, and (b) $u_i$, as a function
    of time for a laminar flow.  The norm of $u_y$ remains small
    throughout the decay. ($\rho_0=10^7$, $v_0=.02$, $D=10$, $n=10$)}
  \label{fig:laminar}
\end{figure}
At the other extreme, at low values of $R$, the flow is laminar.  This
is illustrated in Fig.~\ref{fig:laminar}.  In contrast to the
turbulent case, $\|u_y\|_2$ remains small throughout the run.  ($u_y$
is being continuously driven nonlinearly by $u_x$, so its amplitude
decays with $u_x$.)  With small initial velocities, the linear
analysis of Sec.~\ref{sec:initialdata} is applicable, and the measured
decay rate should be consistent with the prediction
\eqref{eq:shearfrequencies}. Fig.~\ref{fig:lamvort} shows that this is
indeed the case.

\begin{figure}
  \centering
  \begin{tabular}{c c}
    \subfloat[Velocity]{\label{fig:intu}\includegraphics[width=.5\textwidth]{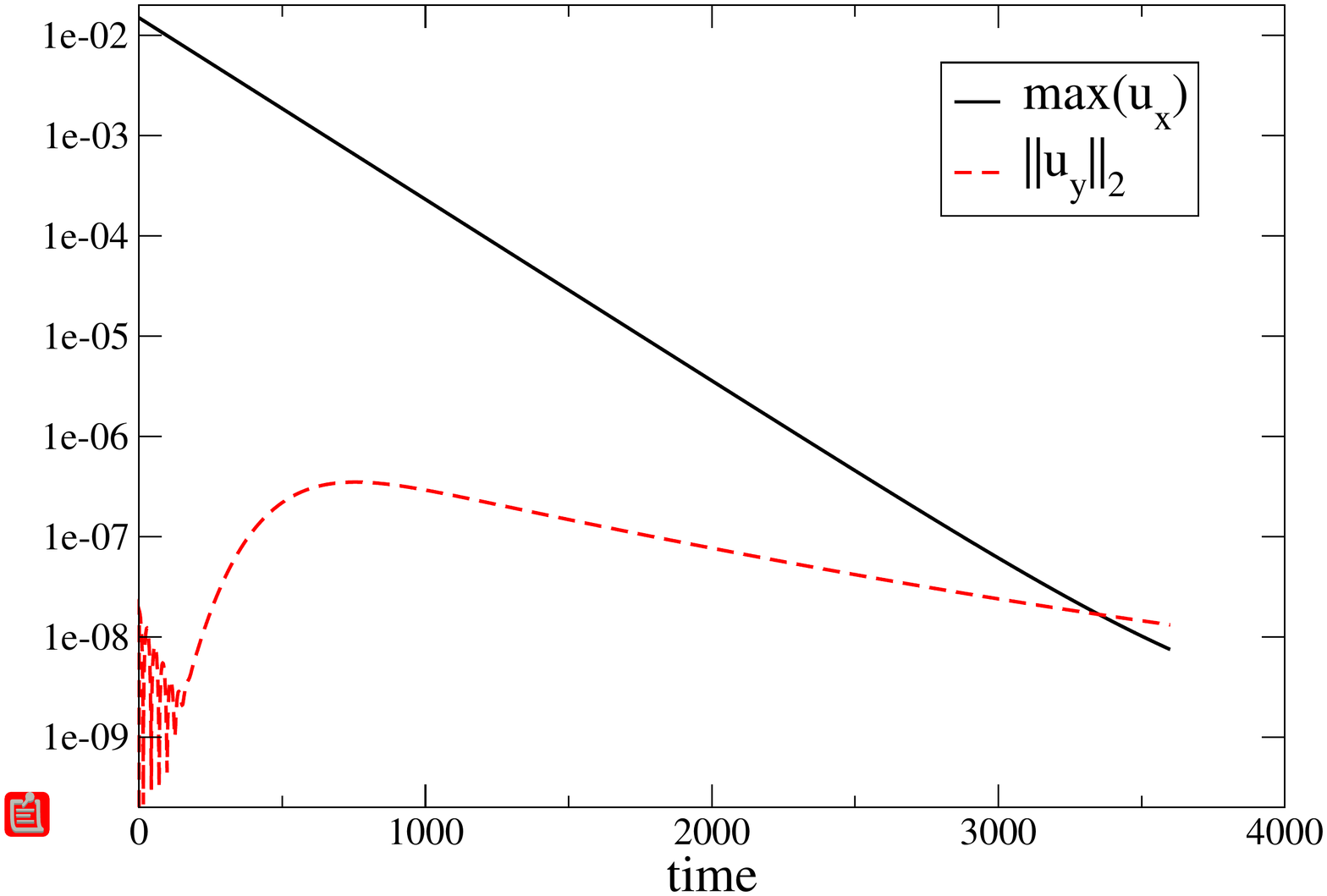}}&
    \subfloat[Growth rate of $\|u_y\|_2$]{\label{fig:growthofuy}\includegraphics[width=.5\textwidth]{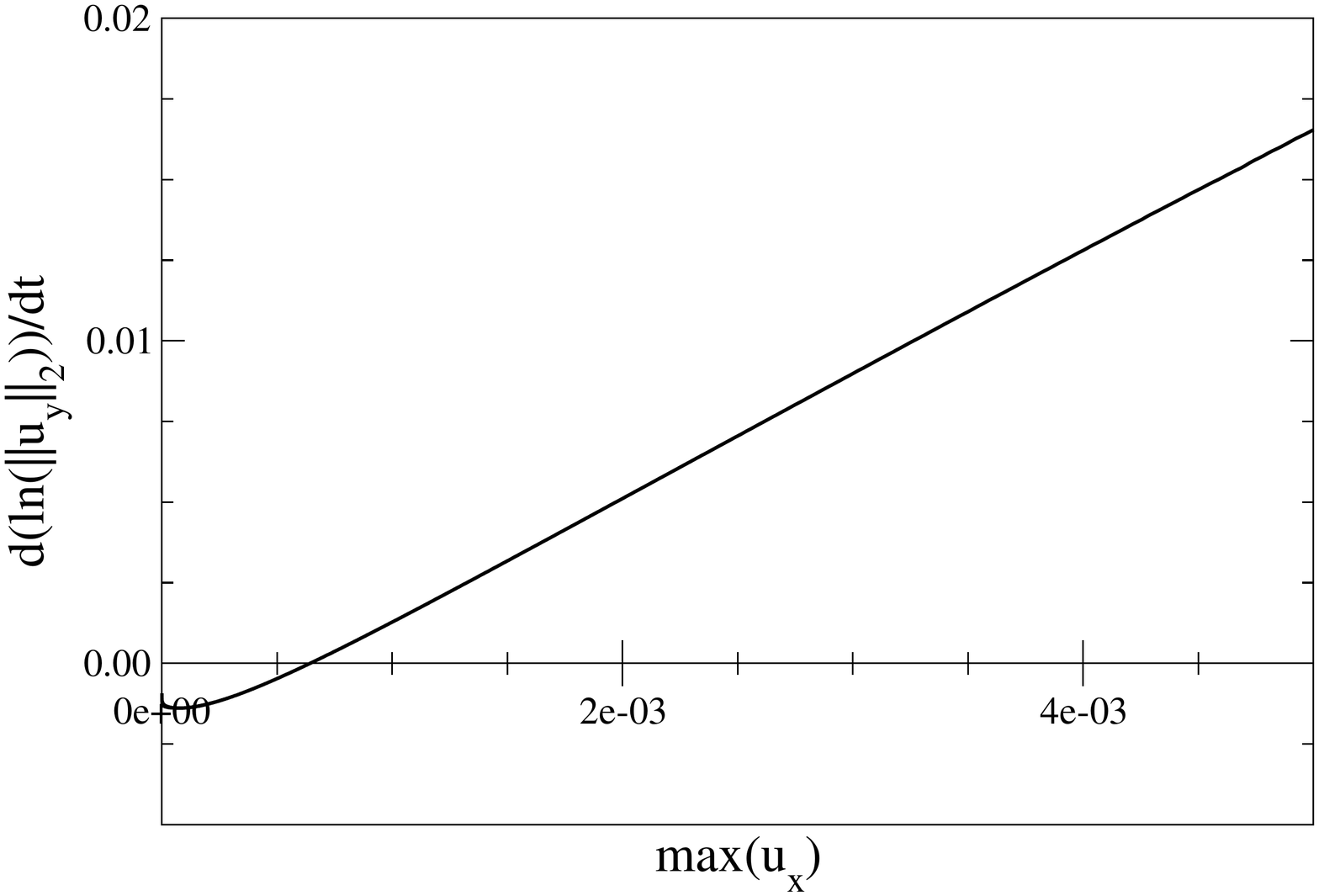}}
  \end{tabular}
  \caption{Intermediate flow: (a) $u_i$, as a function of time for
    laminar flow, and (b) growth rate of $\|u_y\|_2$ versus $\max
    u_x$.  The zero of (b) corresponds to the critical Reynolds
    number.  ($\rho_0=10^7$, $v_0=.015$, $D=4$, $n=10$)}
  \label{fig:intermediate}
\end{figure}
Between these two extremes, we found that there were certain
intermediate flows which provided important physical information.
Such an example is illustrated in Fig.~\ref{fig:intermediate}.  In
this case, the flow begins at high Reynolds number but it decays
before turbulence can fully develop.  The plot of $\|u_y\|_2$ shows
initially small perturbations growing nearly exponentially for some
time -- as in the turbulent case -- before peaking, then decaying
exponentially.

The time dependence of $R$ in such runs is clearly evident, as it is
directly proportional to $\max(u_x)$ (also plotted).  Initially,
$R>R_c$, but as time progresses, $R$ decreases.  Since the background
flow is slowly varying, we assume that the instantaneous growth rate
of $\|u_y\|_2$ depends only on the background value of $\max(u_x)$.
Thus, at the peak of $\|u_y\|_2$, $R=R_c$, while for $R<R_c$,
$\|u_y\|_2$ decays.  This allows us to extract $R_c$.  Indeed in
Fig.~\ref{fig:growthofuy} we plot the growth rate of $\|u_y\|_2$
versus $\max(u_x)$.  Where the curve crosses through zero corresponds
to the peak of $\|u_y\|_2$ in Fig.~\ref{fig:intu}, and the value of
$R_c$ may be read off.  Thus, such intermediate runs provide detailed
information on the stability of the background flow.

\begin{figure}
  \centering
  \includegraphics[width=.5\textwidth]{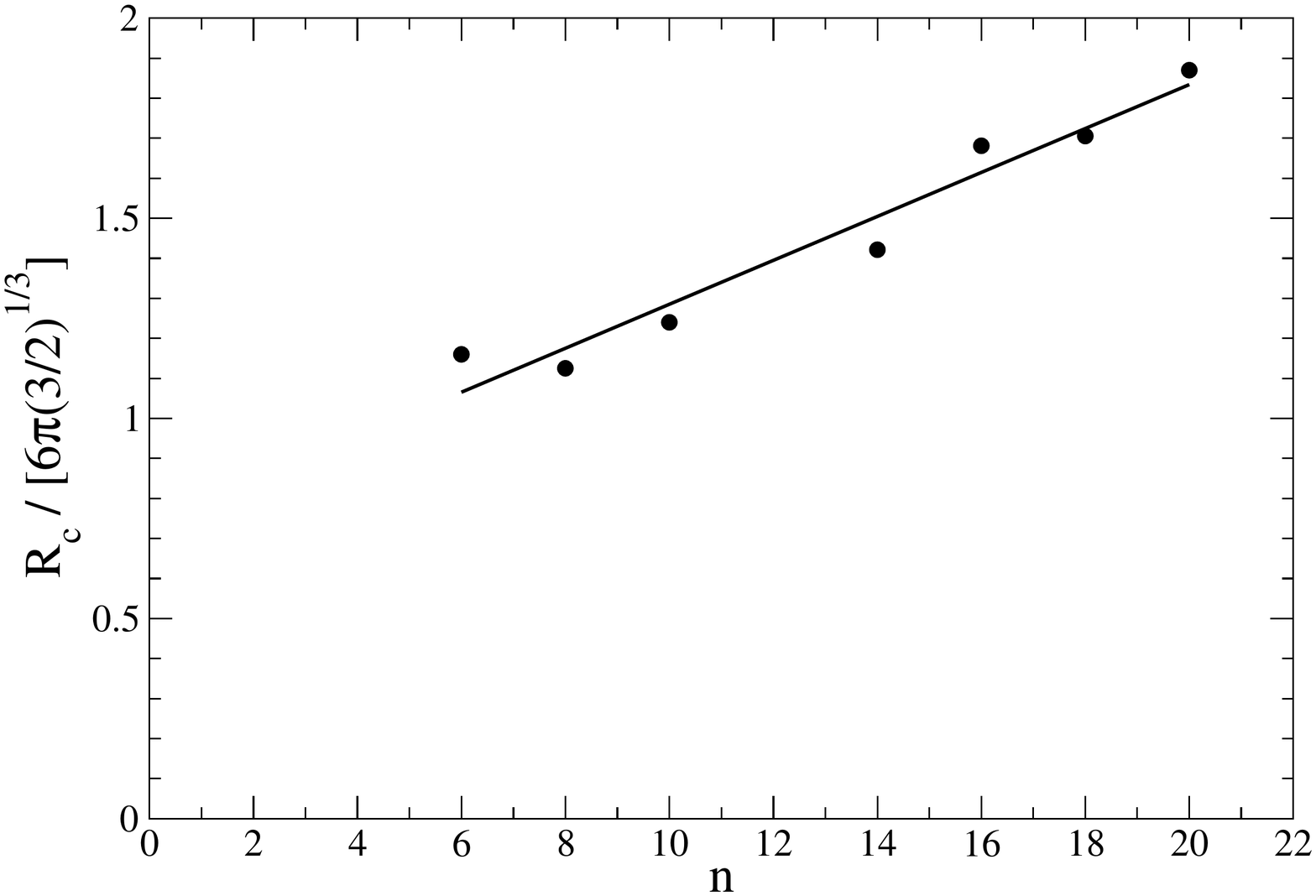}
  \caption{Critical Reynolds number as a function of wave number $n$ (dark circles).
    The solid line is the linear fit~\eqref{eq:Rcfit}.}
  \label{fig:critR}
\end{figure}
By searching for such critical runs (by adjusting $v_0$) for different
values of $(\rho_0,\,D,\,n)$, we found that the critical Reynolds
number of the shear flow is
\begin{equation}\label{eq:Rcfit}
  R_c \approx \left(\frac{3}{2}\right)^{1/3}6\pi\times 0.7 \left(1+0.07n\right).
\end{equation}
The dependence upon $n$ (see also Fig.~\ref{fig:critR}) is due to the
finite size of the box, as discussed above.  We measured this within
the range $6\le n \le 20$.  (We verified, by varying also $\rho_0$ and $D$, that $R_c$ is
independent of these quantities.) For large $n$, we expect $R_c$ to approach
a limit, as the finite box size should play no role.

Fig.~\ref{fig:growthofuy} also provides some information about the
growth rate of $u_y$ away from its zero value at $R=R_c$.  Fixing $n$
and $\rho_0$, this shows that for $R>R_c$, higher $R$ [higher
$\max(u_x)$] gives faster growth.  For $R\gg R_c$, the growth rate is
linear in $\max(u_x)$.  In contrast, as $R$ is lowered, there is a bound on
the decay rate.  This can be understood as a competition between a
driving effect from the background shear flow in $u_x$, and a viscous
decay \eqref{eq:shearfrequencies} associated with the shear mode
\eqref{eq:growinguy} in $u_y$.  Once the driving term drops to the
point of irrelevancy, all that is left is the decay term, which gives
a fixed decay rate, as it is a property of the mode in $u_y$.  Indeed,
the asymptotic value of the growth rate of $\|u_y\|_2$ in
Fig.~\ref{fig:growthofuy} is -0.0013.  This matches very nicely the
prediction from Eq.~\eqref{eq:shearfrequencies} using the observed
$m=6$.

We also found that for large $R$, the growth rate of $\|u_y\|_2$ is
inversely proportional to $\lambda$, at fixed $\max(u_x)$.  Together
with the above, this points to the contribution of the driving term
to the growth rate as being proportional to $\partial_y u_x$.

\subsection{Turbulent regime}

We now turn our attention back to the turbulent case, at the point
where equipartition of energy is reached between $u_y$ and $u_x$
(i.e., just subsequent to Fig.~\ref{fig:turb-b}).  As seen in
Fig.~\ref{fig:turb-c}, the overall flow is completely disrupted, and
the vorticity field displays a number of turbulent eddies.  The
exponential decay of vorticity in Fig.~\ref{fig:turbvort} ceases, and
is replaced with a power law decay.  We typically observed
$\|\omega\|_2\propto t^{-\alpha}$, with $\alpha\simeq 1.2 \pm 0.2$.
There have been several previous studies of unforced turbulent decay
of Navier-Stokes fluids in $d=3$, which have also found power law
decays
(e.g.,~\cite{Matthaeus:1991,PhysRevLett.71.2583,1994PhFl....6.3765H}).

During the power law decay, the eddies merge into vortices, which
continue to merge into increasingly large vortices
(Figs.~\ref{fig:turb-d} and \ref{fig:turb-e}).  This {\em inverse
  energy cascade} can be attributed to the conservation of {\em
  enstrophy}~\cite{Kraichnan:1967}.  Co-rotating vortices merge, while
counter-rotating vorticies repel.  Thus, one is finally left
with two counter-rotating vortices, as in the inviscid
case~\cite{Carrasco:2012nf}.

\subsection{Late time decay}

\begin{figure}
  \centering
  \includegraphics[width=.5\textwidth]{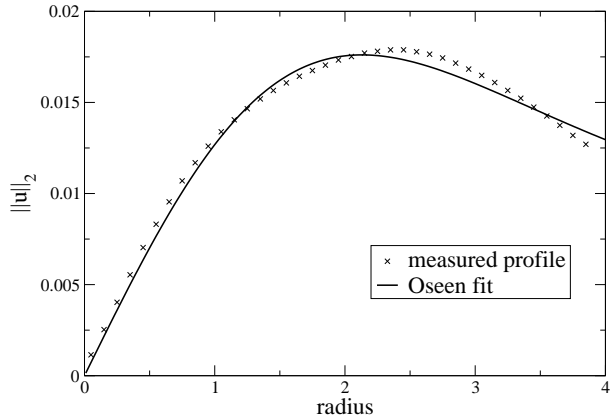}
  \caption{Velocity profile of one of the vortices in
    Fig.~\ref{fig:turb-f}.  The solid line is the fit to the Oseen
    vortex profile.}
  \label{fig:oseen}
\end{figure}

The vortices which form are the relativistic analog of the Oseen
vortex, which is an attractor solution to the Navier-Stokes
equation~\cite{Gallay:2005}.  Its functional form is
\begin{equation}
  v_\theta = \frac{C_1}{r}\left(1-e^{-r^2/C_2(t)^2}\right),
\end{equation}
where the parameter $C_2(t) = \sqrt{4\nu t}$, and the kinematic
viscosity $\nu=\eta/\rho$.  A fit for the parameters $C_1$ and $C_2$
is shown in Fig.~\ref{fig:oseen}.
We see that the vortex is close to, but does not match exactly, the
Oseen profile.  We attribute this to differences betwen Navier-Stokes
fluids, and the relativistic compressible fluids we
study\footnote{Evslin and Krishnan have found exact vortex solutions
  to the relativistic fluids we study here. However as a result of
  imposing stationarity, these solutions are singular, and do not describe
  ours at late times~\cite{Evslin:2010ij}.}.  Fitting also in time, it is possible
to extract $\nu$ from these profiles.  As with the profile fit, we
found a value of $\nu$ which, while not quite the predicted value, was
within a factor of two.

At late times, the two vortices continue to decay in amplitude, until
the fluid finally becomes linear.  The solution is then the sum of
long-wavelength shear and sound modes, which decay exponentially. We
measured the decay rates at late times.  The decay was slightly
faster than the linear prediction, although the measured difference 
decreases at later times.

As a result of the increased wavelength, the decay rate at late times
is lower than the decay rate of the initial flow.  Thus, the presence
of the turbulent cascade drastically lengthens the time period before
the fluid settles down from its initial state to a uniform flow.  In
contrast, for $d>3$, where we expect a direct cascade, turbulence
causes a more rapid decay than the linear behavior. This is due to
higher modes decaying faster and strong dissipation at the viscous scale.

\section{Discussion: Black Branes and Turbulence}\label{sec:final}

In the previous section, we have established conditions for the onset
of turbulent phenomena in conformal, viscous relativistic fluids in
$2+1$ dimensions, as well as the subsequent behavior once this
develops.  Having studied solutions to the dual hydrodynamic theory,
we turn in this section to general relativity, and to the behavior
of perturbed AAdS black holes and black branes.

\subsection{Decay of perturbations and turbulence in the bulk}

The decay properties of the shear fluid flow that we have analyzed in
Sec.~\ref{sec:results} carry over directly to the shear hydrodynamic
quasinormal mode of the black brane.  Indeed, as described in the
introduction, the gravity/fluid correspondence naturally captures
the behavior of the lowest lying shear and sound families of
quasinormal modes.  (As illustrated in Ref.~\cite{Adams:2013vsa}, the
higher order quasinormal modes typically decay very rapidly, and the
metric produced via the duality is a very good approximation to a
solution of Einstein's equation.)

Translating to the black brane language, our results imply that, for
$R>R_c$, hydrodynamic shear quasinormal modes are unstable to small
perturbations (the ``instability'' refers to the quasinormal mode; not to the black brane).  
More precisely, certain deviations from the pure
quasinormal mode undergo exponential growth until {\em either} they
reach the amplitude of the quasinormal mode (fully developed
turbulence) {\em or} the Reynolds number -- which decays in
time -- becomes smaller than $R_c$ and an exponential decay ensues.
Once turbulence sets in, for 4-dimensional AAdS black branes, energy
is transferred to longer wavelength modes and a polynomial decay is induced.
Eventually, when metric deviations about the uniform black brane
become small enough, exponential decay resumes.  In both cases, the final decay is 
at a slower rate than the original decay, as the perturbation is of longer
wavelength.

On the other hand, for $R<R_c$, the quasinormal mode is stable, so it
exhibits the usual clean exponential decay. We stress that in all cases 
described, the global norm of the solution decays in time. 

More generally, one is interested in the behavior of a {\em generic}
black brane perturbation, containing many modes of small but
comparable magnitude.  The standard picture states that if the
amplitude of the perturbation is small enough, then at sufficiently
late times it asymptotically approaches\footnote{Quasinormal modes do
  not in general form a complete basis for solutions, so one cannot
  write an arbitrary solution as a converging sum of modes, except in
  an asymptotic sense.  This was recently demonstrated for
  2-dimensional AAdS black holes~\cite{Warnick:2013hba}.} a sum of
quasinormal modes, which evolve in time independently.  However, the
question is how small the amplitude must be for this to be realized;
this is determined by the Reynolds number.  At high Reynolds number,
certain quasinormal modes are unstable; and an unstable mode will
never be realized in a decay. The new picture that emerges is that of
both {\em laminar} and {\em turbulent} phases.  The laminar phase
corresponds to the standard quasinormal mode picture.  At high
Reynolds number, however, the decaying perturbation immediately enters
the turbulent phase, which has been uncovered through the
gravity/fluid correspondence.

This new turbulent phase displays a far richer phenomenology.  Our
results indicate that turbulence, when it develops, induces eddies.
Eddies with vorticity of the same sign merge, leading to increasingly
large vortices as time
proceeds~\cite{NCAR_OpenSky_OSGC-000-000-015-145}.  The form of the
bulk metric [to leading order, Eq.~\eqref{eq:metric0}], indicates that
the boundary fluid structure is carried unperturbed along ingoing null
geodesic ``tubes''~\cite{Bhattacharyya:2008mz}, connecting the
boundary at spatial infinity to the black brane event horizon.  Thus
the Oseen-like vortices present at late times in the turbulent fluid
solution describe rather compact distributions of gravitational
radiation connecting the asymptotic and black hole regions.  They can
be regarded as natural realizations of ``extended geons'', which
extend through the bulk as ``gravitational tornadoes'' or
``funnels''\footnote{These should not be confused with ``black
  funnels''~\cite{Hubeny:2009kz}, which are bulk black holes with a
  horizon that connects to the conformal boundary of the
  spacetime. Evslin~\cite{Evslin:2012zn} has conjectured these black
  funnels to correspond to singular fluid vortices.}.  The structure
that we describe is illustrated in Fig.~\ref{fig:bulkC2}, where we
plot the curvature invariants $I_1$ and $I_2$ (see
Appendix~\ref{app:geoquant}).
\begin{figure}
  \centering
  \begin{tabular}{c c}
    \subfloat[$r^6\,C_{ABCD}C^{ABCD}$]{\includegraphics[width=.48\textwidth]{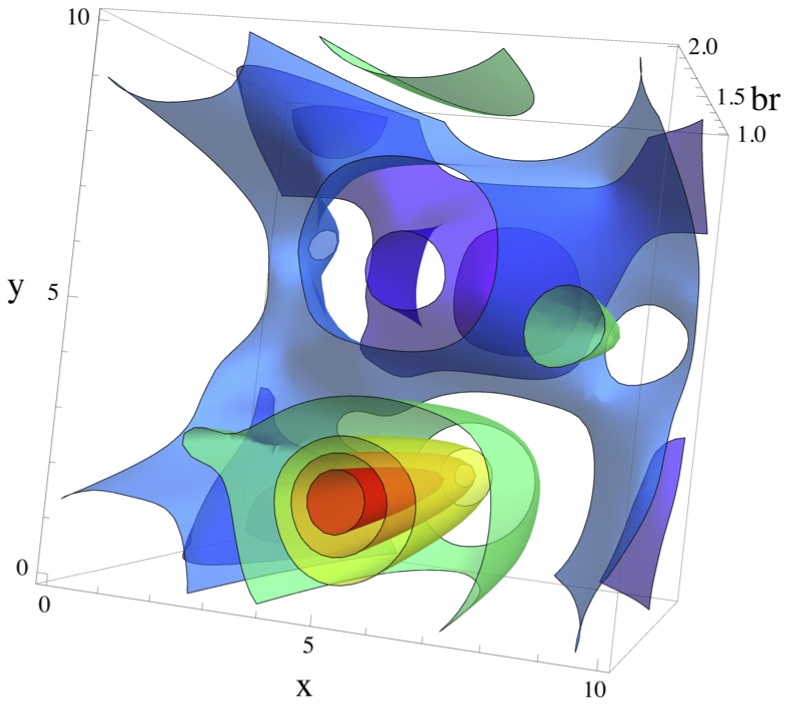}}&
    \subfloat[$r^7\,{}^*C_{ABCD}C^{ABCD}$]{\includegraphics[width=.48\textwidth]{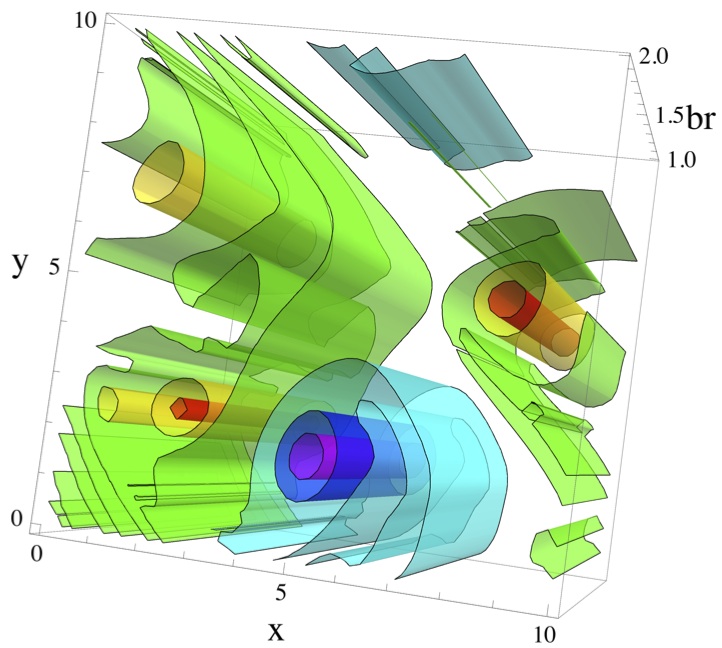}}
  \end{tabular}
  \caption{Contour plots of principal invariants of the Weyl tensor in
    the bulk, computed from the zeroth order metric
    \eqref{eq:metric0}, from the simulation snapshot in
    Fig.~\ref{fig:turb-e}. Notice that (a) is representative of the
    energy density $\rho$, while (b) is representative of the
    vorticity, as expected from Eq.~\eqref{eq:weylinvariants}.  }
  \label{fig:bulkC2}
\end{figure}

As discussed in Appendix~\ref{app:radialmap}, a possible way to
understand this behavior is related to the fact that a map to
relativistic hydrodynamics is also possible away from the boundary, at
suitably defined timelike surfaces in the bulk. Thus, the solution
can also be analyzed at these surfaces to show, in particular, that a
conserved quantity related to the enstrophy can be defined away from
the boundary.  Such a quantity is a key element needed to argue that
an inverse energy cascade occurs.

Naturally, as already pointed out in~\cite{Evslin:2012zn}, it would be
very useful to develop a spacetime definition of the Reynolds number.
This would provide an intrinsic way to predict the onset of turbulence
in gravity and could thus be applied in broader contexts.  Using the
gravity/fluid correspondence, this would also lead to a Reynolds
number suitable for {\em relativistic} hydrodynamics.  Based upon the
form of the bulk metric~\eqref{eq:metric0}, and the fluid Reynolds
number~\eqref{eq:Rnumber}, the form of the Reynolds number for black
hole perturbations is, roughly,
\begin{equation}
  R_{GR} \propto T_{\text{Hawking}}\left\| h_{AB}\left(\frac{\partial}{\partial r}\right)^B\right\| L,
\end{equation}
where we have substituted (certain components of) the metric perturbation
$h_{AB}\equiv g_{AB}-g_{AB}^{(0)}$ for the velocity fluctuation;
and $L$ is the characteristic length scale of the perturbation.  Of
course, whether or not this is applicable in more general contexts
would require further investigation.  In particular, a suitable
definition of $R$ should be gauge invariant.

As a final comment, we point out an important application of the fact
that the inverse cascade guarantees the system stays within the domain
of validity of the gravity/fluid correspondence.  The relativistic
hydrodynamic equations in $2+1$ dimensions are dual to
(long-wavelength) perturbed black branes in the bulk.  Thus, a
Newtonian limit in the bulk -- where time derivatives are taken to be
one-order subleading to space derivatives and velocities are taken to
also be small\footnote{For some alternative discussions, see,
  e.g.,~\cite{Bhattacharyya:2008kq,Bredberg:2011jq}.} -- corresponds to
a Navier-Stokes limit on the boundary.  Since we know that the
Navier-Stokes equation admits global, well-behaved
solutions~\cite{La69}, one can surmise that general relativity
is similarly well-behaved in the bulk.


\subsection{Connection to ordinary perturbation theory}

Linear perturbation theory predicts that small-amplitude metric
perturbations can be decomposed (asymptotically) into independent modes which undergo
simple exponential decay.  On the other hand, the picture described in
the previous section indicates the presence of a qualitatively
distinct, turbulent, behavior for sufficiently high Reynolds numbers,
regardless of the perturbation amplitude. How are these two notions
reconciled?  The short answer is that at very high black hole
temperatures (in AdS), the regime of applicability of linear
perturbation theory is very small.  To study this further, we take a
closer look at perturbation theory.

In ordinary perturbation theory, the full metric is expanded as
\begin{equation}
  g_{AB} = g_{AB}^{(0)} + h_{AB}^{(1)} + h_{AB}^{(2)} + \ldots,
\end{equation}
where $g_{AB}^{(0)}$ is taken to be the background metric; in our
case, the uniform AdS black brane.  The first order metric
perturbation $h_{AB}^{(1)}$ satisfies the homogeneous partial
differential equation,
\begin{equation}\label{eq:linEinstein}
  G_{AB}^{(1)}(g^{(0)},h^{(1)}) + \Lambda h_{AB}^{(1)} = 0,
\end{equation}
where $G_{AB}^{(1)}(g^{(0)},h^{(1)})$ is the linearized Einstein
tensor.  For the black brane, the symmetry properties admit a mode
decomposition, and by solving Eq.~\eqref{eq:linEinstein}, the
quasinormal mode spectrum is determined.  All of the modes decay for
the AdS black brane \cite{Berti:2009kk,Morgan:2009pn}.

At second order in perturbation theory,
\begin{equation}
  G_{AB}^{(1)}(g^{(0)},h^{(2)}) + \Lambda h_{AB}^{(2)} = - G^{(2)}_{AB}(g^{(0)},h^{(1)}),
\end{equation}
where the second order Einstein tensor on the right hand side is
quadratic in the first order perturbation $h^{(1)}_{AB}$.  Since the
homogeneous part of this equation is unchanged from the first order
case, the quasinormal mode spectrum of $h_{AB}^{(2)}$ is also
unchanged.  These decaying modes are excited by the inhomogeneous
source term.  At any finite order in perturbation theory, the same
applies.  Thus, the growth we describe can be only captured by carrying out
the analysis to sufficiently high orders in perturbation theory to recognize
the underlying exponential behavior.

If, rather than taking the background metric $g_{AB}^{(0)}$ to be the
uniform AdS black brane, it is instead taken to be the AdS black brane
{\em plus the shear hydrodynamic quasinormal mode}, then the growth is
easily seen to be possible at the linearized
level~\eqref{eq:linEinstein}. This is best illustrated through a simple toy
model of ordinary differential equations (inspired by a local analysis of the Navier-Stokes
equations) which exhibits similar mode-coupling
properties, namely 
\begin{align}
  \label{eq:x}\frac{dx}{dt}+\alpha x &= 0,\\
  \label{eq:y}\frac{dy}{dt}+\beta y - \gamma x y&= 0,
\end{align}
with $\alpha$, $\beta$, and $\gamma$ all positive constants.  The
variable $x(t)$ in this system corresponds to $u_y$ in the black brane
system, which initially describes a shear mode.  On the other hand,
$y(t)$ corresponds to the initially zero $u_y$.  Both $x$ and $y$ are
subject to dissipation (in the black brane case, $\alpha \approx
\beta$), and we have included a mode-coupling $\gamma x y$ in
\eqref{eq:y}.

Let us solve the system \eqref{eq:x}--\eqref{eq:y} perturbatively in
two different ways\footnote{We note that Eqs.~\eqref{eq:x}--\eqref{eq:y} can easily be solved exactly.  Nevertheless, the perturbative methods described here carry over directly to more complicated systems.}.  We expand both $x$ and $y$ as
\begin{align}
  x &= x_{(0)} + x_{(1)} + x_{(2)} + \ldots,\\
  y &= y_{(0)} + y_{(1)} + y_{(2)} + \ldots.
\end{align}
The exact solution to~\eqref{eq:x} is $x \propto e^{-\alpha t}$.  In a
manner analogous to perturbation theory about the {\em uniform} black
brane, we take the ``background solution'' to be $x_{(0)} = y_{(0)} =
0$, while the ``linearized solution'' corresponds to the quasinormal
mode, i.e., $x_{(1)}= a_1 e^{-\alpha t}$.  Then, the equation for
$y_{(1)}$ is
\begin{equation}\label{eq:toy1}
  \frac{dy_{(1)}}{dt} + \beta y_{(1)} = 0,
\end{equation}
with solution $y_{(1)}= b_1 e^{-\beta t}$.  At second order,
\begin{equation}
  \frac{dy_{(2)}}{dt} + \beta y_{(2)} = \gamma x_{(1)} y_{(1)} = \gamma a_1 b_1 e^{-(\alpha + \beta)t}.
\end{equation}
The solution to this is $y_{(2)} = -a_1b_1\gamma
e^{-(\alpha+\beta)t}/\alpha + b_2 e^{-\beta t}$; higher order corrections can be
computed in a similar manner.  At each
order in perturbation theory the solution is a sum of decaying
exponentials.

Suppose instead that we take $x_{(0)} = a_0 e^{-\alpha t}$ and
$y_{(0)} = 0$.  This is analogous to taking the black brane {\em
  perturbed by a quasinormal mode} as the ``background solution''.
Then, at linear order we have
\begin{equation}\label{eq:toy2}
  \frac{dy_{(1)}}{dt} + \beta y_{(1)} = \gamma x_{(0)} y_{(1)} = \gamma a_0 e^{-\alpha t} y_{(1)},
\end{equation}
which has solution 
$y_{(1)} =
b_1 \exp\left[-\beta t + a_0\gamma\left(1-e^{-\alpha
      t}\right)/\alpha\right]$.  In the limit of small $\alpha$, this
becomes
\begin{equation}
  y_{(1)}\xrightarrow{\alpha \to 0}b_1 e^{-(\beta - a_0\gamma) t}.
\end{equation}
Thus, if the ``background'' is long-lived, the growth rate of $y$
depends upon a competition between the driving term $\gamma x y$
and the dissipative
term $\beta y$.  For finite $\alpha > 0$, the driving term decreases
with time relative to the dissipative term.  In this case,
$y_{(1)}$ is eventually dominated by the exponential decay
(cf.~Fig.~\ref{fig:intermediate}).

One can define a ``Reynolds number'' of the flow $x(t)$ in the toy
model, by taking the ratio of the mode-coupling term $\gamma x y$, to
the linear dissipative term, $\beta y$, in \eqref{eq:y}.  If this
``Reynolds number'', $\gamma a_0 e^{-\alpha t}/\beta$, is large, then
the nonlinear term $\gamma x y$ should be kept in any perturbative
analysis, and one should solve Eq.~\eqref{eq:toy2} -- rather than
Eq.~\eqref{eq:toy1} -- to determine $y$.  

A similar story holds in for black brane perturbations in general
relativity.  The problem with trying to analyze high Reynolds number
perturbations by performing ordinary perturbation theory about the
uniform black brane background, is that this drops certain large
nonlinear terms while keeping small linear terms.  For the black
brane, the dissipation rate (the analog of $\alpha \approx \beta$)
depends inversely on the temperature. So, increasing the temperature,
while holding fixed the amplitude and wavelength of a perturbation
(i.e., increasing the Reynolds number), the linear term evantually
becomes small relative to nonlinear terms.  Thus, the regime of
validity of ordinary perturbation theory is reduced as the Reynolds
number is increased.

\subsection{Beyond 4-d AAdS black branes}

Here we discuss some possible implications and extensions of this work.

\subsubsection{Higher dimensions}
Based upon our results for 4-dimensional bulk spacetimes,
together with the gravity/fluid correspondence established in
arbitrary dimensions, as well as numerical results confirming the
expectation of direct energy cascades for inviscid conformal relativistic
fluids~\cite{Radice:2012pq}, it is possible to anticipate properties
of 5 (and higher) dimensional spacetimes (for an early
discussion, see~\cite{VanRaamsdonk:2008fp}).  Three immediate
consequences are:
\begin{itemize}
\item First, as with the 4-dimensional case, at sufficiently high
  Reynolds number, the quasinormal mode description fails to
  accurately describe the decay of black brane perturbations.
\item Second, in contrast to the 4-dimensional case, turbulence is
  characterized by a {\em direct energy cascade} to short wavelengths.
  Since shorter wavelengths have a more rapid decay rate,
  thermalization in 5 dimensions will be attained in a shorter
  timescale than in an analog 4-dimensional case.  Notice that a
  potential concern here is that the cascade to short wavelengths may
  cause the solution to exit the regime of validity of the
  gravity/fluid correspondence.  Two comments are relevant here.  (i)
  Even if this were the case, perturbations initially satisfying $LT
  \gg 1$ (as required by the correspondence) still undergo turbulent
  dynamics which induce structure at shorter wavelengths. Thus, at the
  moment of ``exiting'' the regime of validity, many modes would be
  present and their subsequent behavior should be studied within
  general relativity.  (ii) It is also possible for the cascade to
  occur completely within the regime of validity of the correspondence
  if the viscous scale (again defined as in the non-relativistic case)
  $L_{\eta} = (\eta^3/\epsilon)^{1/4}$ (where $\epsilon$ is the rate
  of energy dissipation by viscosity) satisfies $L_{\eta} T \gg
  1$. Notice that $L_{\eta}$ grows with temperature, so at
  sufficiently high $T$, this condition is satisfied.  In this case
  energy would be expected to cascade down to the viscous scale, and
  then dissipate. 

  As a consequence of the turbulent behavior on the hydrodynamical
  side -- which displays a self-similar behavior -- one would expect a
  similar (fractal) structure for black hole perturbations on the
  gravitational side\footnote{This was very recently observed for
    4-dimensional black holes~\cite{Adams:2013vsa}.  Black holes with
    a fractal structure have also been recognized for a class of
    unstable black holes in higher dimensions~\cite{Lehner:2010pn},
    although any connection to turbulence is yet to be understood.}.
  This structure is expected to smooth out in time yielding a slightly
  hotter, uniform black brane.  If this is the case, then in this high
  temperature limit [case (ii)], a global solution to the dual
  relativistic hydrodynamics problem would seem to be a natural
  consequence.  As in the $d=2+1$ fluid case, this would have obvious
  implications for global solutions to the Navier-Stokes equation.
  Establishing this decay result rigorously on the gravity side
  amounts to determining nonlinear stability\footnote{Related work in
    the linear case has been presented
    in~\cite{Holzegel:2013kna,Holzegel:2011uu}.} of large black holes
  in AdS.

\item Finally, a word of caution with regard to nonlinear numerical
  studies of gravitational perturbations in 5 dimensions: Due to
  the high computational cost of such simulations, symmetries are
  usually imposed on the solution, effectively reducing the number of
  dimensions which are actually simulated.  However, this may restrict
  or affect the development of turbulence. In particular,
  5-dimensional spacetimes which are dimensionally reduced to
  4 dimensions will give rise to an inverse energy
  cascade -- rather than a direct one -- while 3-dimensional
  treatments will eliminate turbulence altogether.  As one is often
  interested in using the AdS/CFT correspondence to describe high
  temperature CFTs through a gravitational analysis, it is important
  to bear in mind that the imposition of computation-saving symmetries can
impact the extracted physics.
\end{itemize}

\subsubsection{Black holes}

We have investigated the decay of black branes in a Poincar\'e patch
of AdS, with torus topology.  However, it is also of interest to study
black holes in global AdS.  Based on earlier work in the inviscid
case~\cite{Carrasco:2012nf}, we expect a qualitatively similar
turbulent behavior. One primary difference, though, is the final
number of vortices which remain (e.g., fluids dual to
Kerr-Schwarzschild settle down to two clockwise and two
counter-clockwise rotating vortices, while Kerr-AdS settles to just
one of each sign).  Other particular details, such as the power law
decay rate during turbulence, and the critical Reynolds number, may
also differ. The power law behavior, $\|\omega\|_2\propto t^{-\alpha}$,
can also be estimated in the inviscid case.  An examination of
results presented in~\cite{Carrasco:2012nf} indicates that for black
holes, the decaying behavior is realized with a similar exponent to
black branes, in the range $0.5 < \alpha < 1.5$.

\subsubsection{Beyond AdS}

At a speculative but certainly tantalizing level, that gravity
displays turbulent behavior in AAdS spacetimes suggests that more
general asymptotic conditions should also be investigated.  Is this a
special property of AdS, or could it arise in the asymptotically flat
or de Sitter cases?  What about Dirichlet boundary conditions, but
without a cosmological constant?  There are two elements to consider:
The boundary conditions imposed on the solutions to the Einstein
equation, and the presence of the cosmological constant in the
equation of motion.  From a partial differential equations point of
view the cosmological constant introduces a lower order term in the
equations, which does not affect local propagation.

On the other hand, it is well known that linearized perturbations of
AAdS black hole spacetimes have a family of very slowly decaying modes
(the hydrodynamic modes), which (as we have shown) play a key role in
terms of being unstable to perturbations.  Such modes are also present
in certain vacuum solutions bounded by an accelerating mirror, which
have been shown to be dual to Navier-Stokes
fluids~\cite{Bredberg:2011jq}.  Interestingly, at least some
hydrodynamic modes can also be connected to QNMs of asymptotically
flat spacetimes~\cite{santos}. However, in this regime, the modes
decay rapidly, and therefore do not govern the long term behavior of
the system.  However, they could play a role in channeling energy in
the transient stages.  Furthermore, it is well known that massive
fields introduce {\em effective} boundaries which could induce the
long-lived mode behavior, and thus allow for turbulent-like phenomena.  As
discussed earlier, if a Reynolds number can be suitably defined for
gravity, it would help predict the onset of turbulence in these varied
scenarios.

\subsection{Final words}

As we have stressed throughout this work, the gravity/fluid
correspondence translates intuition of fluid behavior into the realm
of gravity.  This has allowed us to identify key features of the
behavior of perturbed AAdS black holes, including a new dynamical
phase where fully developed ``turbulence'' gives rise to a polynomial
decay. Furthermore, turbulent behavior also indicates that perturbed
black holes can behave in a strongly dimensionally-dependent way, both
qualitatively and quantitatively. Establishing that gravity can behave
in a turbulent manner opens new doors to searching for other
situations where this can take place. For instance, it provides
motivation to look for scenarios where slowly decaying perturbations
might give rise to interesting non-linear interactions.  Finally,
insights from turbulence may shed new light on particular systems
known to exhibit related behavior, such as the chaotic behavior of
spacelike singularities in early-universe mixmaster
dynamics~\cite{Evslin:2012zn}.

\begin{acknowledgments}
  We wish to thank A.~Buchel, V.~Cardoso, P. Chesler, M. Krucenszky,
  D. Marolf, R. Myers, R.~Wald and T. Wiseman for helpful discussions.
  This work was supported by NSERC through Discovery Grants and CIFAR
  (to L.~L.)  and by CONICET, FONCYT, and SeCyT-Univ.~Nacional de
  Cordoba Grants (to F.~C.). S.~R.~G. is supported by a CITA National
  Fellowship at the University of Guelph. F.~C. and S.~R.~G. thank the
  Perimeter Institute for hospitality.  This research was supported in
  part by Perimeter Institute for Theoretical Physics. Research at
  Perimeter Institute is supported by the Government of Canada through
  Industry Canada and by the Province of Ontario through the Ministry
  of Research and Innovation. Computations were performed at SciNet.
\end{acknowledgments}

\appendix

\section{Spacetime Turbulence: Bulk Behavior and Radial Map to
  Relativistic Hydrodynamics}
\label{app:radialmap}

As mentioned, the gravity/fluid correspondence indicates that the
boundary behavior is manifested throughout the bulk. Consequently, one
expects an inverse cascade behavior which mirrors the behavior at the
boundary.  Here we provide further details on how this behavior can be
seen to arise. In particular, we show how a conserved current (in the
high temperature limit that approaches the inviscid case) gives rise
to a conserved enstrophy.

\subsection{Preliminary considerations: Boundary quantities and enstrophy}

To first order, and now with an arbitrary boundary metric $\gamma_{\mu \nu}$,
the bulk metric can be written in the form \cite{Bhattacharyya:2008mz}
\begin{equation}
 ds^2_{[0+1]} = -2 u_\mu dx^\mu (dr + r {\cal A}_{\nu} dx^{\nu} ) + 
 r^2 \left( \gamma_{\mu\nu} + \frac{1}{(b r)^d} u_\mu u_\nu + 2 b F(br) \sigma_{\mu\nu} \right) dx^\mu dx^\nu.
\end{equation}
Here, $\mathcal{A}_{\nu} \equiv a_{\nu} - \frac{1}{2}\Theta u_{\nu} $,
where $\Theta \equiv \nabla_\mu u^\mu$ is the expansion of the velocity field $u^\mu$, and $a^{\mu}$ its acceleration.

At the AdS boundary, the following quantities can be defined:
\begin{itemize}
\item The stress-energy tensor \cite{Balasubramanian:1999re,prescription-B},
\begin{equation}
 T^{\mu}_{\phantom{\mu}\nu} \equiv \lim_{r \to \infty} \frac{r^d}{8\pi G_{d+1}} (K_{\phantom{\mu}\nu}^{\mu} - \delta^{\mu}_{\nu} K),
\label{prescription}
\end{equation}
where $K_{\phantom{\mu}\nu}^{\mu}$ is the extrinsic curvature of a
constant-$r$ hypersurface. To
first order in the derivative expansion, this is
\begin{equation}
 T_{\mu \nu}^{[0+1]} =  (\rho + P) u_{\mu} u_{\nu} + P \gamma_{\mu \nu} - 2 \eta \sigma_{\mu \nu} .
\end{equation}
\item Conserved currents. In the high temperature limit, the effects
  of viscosity are subleading. Thus, if $\gamma_{\mu\nu}$ admits a
  timelike Killing vector field $\xi^{\mu}$, then in this limit,
  conservation of stress-energy gives rise to conservation of the
  energy current~\cite{Carrasco:2012nf},
\begin{equation}
 J^{\mu}_{\rho} \equiv \frac{1}{2} \rho (\gamma^{\mu\nu} + 3 u^{\mu} u^{\nu}) \xi_{\nu}.
\end{equation}
Again in the inviscid limit, there is also a conserved enstrophy current \cite{ Carrasco:2012nf},
\begin{equation}
 J^{\mu}_{Z} \equiv  \omega^{\alpha \beta} \omega_{\alpha \beta} u^{\mu}.
\end{equation}
\end{itemize}

\subsection{Bulk behavior: Radial map and enstrophy}
\label{app:Z}

The steps above can be extended into the bulk by considering
$r=\text{constant}$ timelike hypersurfaces\footnote{For a related
  treatment for the null hypersurface corresponding to the horizon
  see~\cite{Eling:2013sna}.}.  Within the regime of validity of the
gradient expansion, such hypersurfaces are timelike outside the
black brane horizon.  


It is possible to project the Einstein equation onto a constant-$r$
hypersurface, in order to obtain a fluid description
on the hypersurface~\cite{Brattan:2011my,Kuperstein:2011fn}.  As in the
$r\to\infty$ limit addressed earlier, the ``momentum constraint''
gives rise to a conserved stress-energy tensor $\hat{T}_{\mu\nu}$.  It
turns out to be of the same form as the boundary stress-energy,
with
\begin{equation}
 \hat{T}_{\mu \nu} =  (\hat{\rho} + \hat{P}) \hat{u}_{\mu} \hat{u}_{\nu} + \hat{P} \hat{\gamma}_{\mu \nu} - 2 \eta \hat{\sigma}_{\mu \nu}+\ldots .
\end{equation}

These new (hatted) fields $\hat{\gamma}_{\mu \nu}$, $\hat{p}$, $\hat{u}_{\mu}$, $\hat{\rho}$ and $\hat{\sigma}_{\mu \nu}$ are related
to the original fields at the boundary through a map, which, to first order in derivatives, gives~\cite{ Brattan:2011my, Kuperstein:2011fn}
\begin{align}
 \hat{\gamma}_{\mu \nu} &= \gamma_{\mu \nu} + \left(1-\frac{1}{\alpha^{2}}\right) u_{\mu} u_{\nu} + 2b F(br) \sigma_{\mu \nu} - \frac{2}{r} u_{(\mu}\mathcal{A}_{\nu)}  + \ldots , 
\label{h-hat} \\
 \hat{\gamma}^{\mu \nu} &= \gamma^{\mu \nu} + \left(1- \alpha^{2} - \frac{\alpha^4 \Theta}{r}\right) u^{\mu} u^{\nu} - 2b F(br) \sigma^{\mu \nu} + \frac{2 \alpha^2}{r} u^{(\mu}a^{\nu)}  + \ldots,\\
 \hat{u}_{\mu} &= \left( 1 - \frac{\alpha^{2} \Theta}{2r}\right) \frac{u_{\mu}}{\alpha} + \frac{\alpha}{r} a_{\mu} ,\\
 \hat{u}^{\mu} &= \left( 1 + \frac{\alpha^{2} \Theta}{2r}\right) \alpha u^{\mu} + \ldots ,
\label{u-hat}\\
\hat{\rho} &= \frac{2\alpha}{\alpha+1} \rho ,
\label{density}\\
\hat{P} &= \frac{\alpha}{\alpha+1} (3\alpha-1) P ,\\
\hat{\omega}_{\mu \nu} &= \frac{1}{\alpha} \omega_{\mu \nu} ,
\label{vor}
\end{align}
where $\alpha \equiv \left(1-\frac{1}{(br)^{3}}\right)^{-1/2} = \left(1-\frac{8\pi G_{4}}{r^{3}} \rho\right)^{-1/2}$.

A crucial difference with respect to the boundary fluid is that now the fluid ``lives'' on a 
background $\hat{\gamma}_{\mu \nu}$, which is dynamical. In addition, this fluid
obeys a more complicated equation of state, 
\begin{equation}
 \hat{P} = \frac{(3\alpha(\hat{\rho})-1)}{4} \hat{\rho},
 \label{state-r}
\end{equation}
where $\alpha(\hat{\rho})=\left(1-\frac{4\pi G_{4}}{r^{3}} \hat{\rho} \right)^{-1} $. Connected 
with this last point, the stress-tensor is no longer traceless,
\begin{equation}
 \hat{T}_{\phantom{\mu}\mu}^{\mu} = \frac{3\alpha(\alpha - 1)}{(\alpha + 1)}  \rho = -r \frac{d\hat{\rho}}{dr} .
\end{equation}
This has been interpreted as an RG flow \cite{Brattan:2011my,Kuperstein:2011fn,Kuperstein:2013hqa}, 
in which the radius $r$ plays the role of an
energy scale from the field theory perspective.

Keeping in mind these differences, it is still the case that each
constant-$r$ timelike hypersurface contains a relativistic fluid
description, just like the AdS boundary. Thus, similar conservation
laws can be derived. Of particular interest is the existence of an
enstrophy which, in the high temperature limit, is conserved.  To
define this quantity for a general equation of
state $\hat{P}(\hat{\rho})$, only a
slight generalization of the previous derivation is required. 

We begin with the inviscid fluid equations, 
\begin{align}
 u^{\nu} \partial_{\nu} \rho  &= -(\rho + P) \nabla_{\mu} u^{\mu} \label{along} ,\\
 P^{\mu \nu} \partial_{\nu} P &= -(\rho + P) u^{\nu} \nabla_{\nu} u^{\mu} \label{orthogonal}  .
\end{align}
Next, we will require a function $\tilde{\rho}(\rho)$ such that the 2-form $\Omega_{\mu \nu}\equiv2\nabla_{[\mu}
(\tilde{\rho} u_{\nu]})$ satisfies $\Omega_{\mu \nu}u^{\nu}=0$.  This is accomplished by choosing
\begin{equation}
 \tilde{\rho}(\rho) \propto (\rho + P) \exp\left[ -\int \frac{d\rho'}{\rho' + P(\rho')} \right],
\end{equation}
where the exponent contains an indefinite integral, and we assumed the integration can be performed 
for a given equation of state $P(\rho)$.

Notice that  $\Omega_{\mu \nu}$, is \textit{strongly conserved} by the flow by virtue of Cartan's identity. That is,
\begin{equation}
 \textbf{\pounds}_{\lambda \mathbf{u}} \mathbf{\Omega} = \lambda \mathbf{u}\cdot \mathbf{d\Omega} + \mathbf{d}(\lambda \mathbf{u}\cdot \mathbf{\Omega}) = 0  .
\label{Lie}
\end{equation}
Motivated by this observation, we can construct a current of the form $J_Z^{\mu}\equiv \lambda \Omega^{2} u^{\mu}$, where we have 
denoted $\Omega_{\alpha \beta} \Omega^{\alpha \beta} \equiv \Omega^2 $,
and $\lambda$ a scalar field to be fixed by the conservation requirement. Computing the divergence of $J^{\mu}$ one obtains,
\begin{align}\label{conservedcurrent}
 \nabla_{\mu}J_Z^{\mu} &= \Omega^2 \left[ \lambda \nabla_{\mu}u^{\mu} + u^{\mu}\partial_{\mu} \lambda \right]  + 2\lambda \Omega^{\sigma \beta} u^{\mu}\partial_{\mu}\Omega_{\sigma \beta}    \nonumber\\
                     &= \Omega^2 \left[ \lambda \nabla_{\mu}u^{\mu} + u^{\mu}\partial_{\mu} \lambda \right]  - 4\lambda \Omega^{\sigma \beta} \Omega_{\sigma \mu} (\nabla_{\beta}u^{\mu})    \nonumber\\
		     &= \Omega^2 \left[ u^{\mu}\partial_{\mu} \lambda - \lambda \nabla_{\mu}u^{\mu} \right]   .
\end{align}
We have used the definition of the Lie derivative and \eqref{Lie} on the second line, and the third line follows from,
\begin{equation}
 \Omega^{\mu \alpha} \Omega_{\mu \beta} = \frac{1}{2} \Omega^{2} P^{\alpha}_{\phantom{\alpha}\beta} ,
\label{key-step}
\end{equation}
an identity which is valid in two spatial dimensions~\cite{Carrasco:2012nf}. 
Clearly, the divergence in Eq.~\eqref{conservedcurrent} vanishes if
$\lambda (\rho) \propto  \exp\left[ -\int \frac{d\rho'}{\rho' + P(\rho')} \right]  $.\\

Notice also that $ \Omega^2 $ is related to the square vorticity ($
\omega^2 \equiv \omega_{\mu \nu}\omega^{\mu \nu} $) by $ \Omega^2 = 4
\tilde{\rho}^2 \omega^2 $.  Finally, by integrating the current
$J_Z^\mu$ over a constant-$t$ hypersurface $\Sigma_t$, the expression
for the enstrophy takes the form
\begin{equation}
 Z \equiv \int_{\Sigma_{t}} \gamma (\rho + P)^2  \exp\left[ -3 \int \frac{d\rho'}{\rho' + P(\rho')} \right] \omega^2  dS^2.
 \label{enstrophy}
\end{equation} 

It is straightforward to check that this expression reduces to the one reported previously in \cite{Carrasco:2012nf} for the conformal fluid on the boundary,
\begin{equation}
 Z_{\text{boundary}} = \int_{\Sigma_{t}} \gamma \omega^2  dS^2 \, ,
\end{equation} 
when the equation of state is given by $P(\rho)=\frac{1}{2} \rho$.\\

We are now in a position to evaluate the enstrophy on an arbitrary $r=\text{constant}$ timelike hypersurface in terms of boundary variables,
\begin{equation}
 Z(r) = \int_{\Sigma_{t}(r)} \hat{\omega}^2 (\hat{\rho} + \hat{P})^2  \exp\left[ -3 \int 
 \frac{d \hat{\rho}'}{\hat{\rho}' + \hat{P}(\hat{\rho}')}\right] \hat{u}^{\mu} d\hat{\Sigma}_{\mu}  .
 \label{enstrophy-r}
\end{equation} 
From Eqs.~\eqref{state-r} and \eqref{density} we get,
\begin{equation}
(\hat{\rho} + \hat{P})^2  \exp\left[ -3 \int \frac{d \hat{\rho}'}{\hat{\rho}' + \hat{P}(\hat{\rho}')}\right] 
= \left( \frac{3}{8}\right) ^2 \alpha^2  ,
\end{equation}
while for the square vorticity we find that,
\begin{equation}
 \hat{\omega}^2 = \frac{1}{\alpha^2} \omega^2 .
\end{equation}

As a final remark, let us note that to zeroth order in the gradient
expansion, the transformation from one surface element to the other is
given by
\begin{equation}
 \hat{u}^{\mu} d\hat{\Sigma}_{\mu} = u^{\mu} d\Sigma_{\mu} .
\end{equation}
Consequently, the enstrophy calculated at any given radius $r$, not
only is conserved, but is also equivalent (up to an unimportant constant factor which can
be trivially absorbed into the definition) to the enstrophy of the
boundary fluid. That is,
\begin{equation}
 Z(r) \sim Z_{\text{boundary}}  .
\end{equation}

The fact that it is possible to define an enstrophy for the fluid on
each slice of the bulk geometry, and that it has the same expression
in all cases, is a natural consequence of the {\em ultra-local}
character of the map.  Thus, in a sense, the dynamics occurring at the
boundary are reproduced throughout the bulk geometry (in a slightly
distorted manner).  At first sight, this enstrophy construction in the
bulk seems to indicate that one could define an interesting local
quantity from the spacetime perspective in the bulk. However, since
this quantity is defined on a distorted (and dynamical) surface, one
should exercise caution, especially in highly distorted scenarios.

\section{Bulk Behavior: Geometrical Quantities}
\label{app:geoquant}

It is also instructive to examine the bulk through
geometrical quantities. This is particularly appealing as it could
provide a geometrical way to understand fluid phenomena. Such a
program, however, is delicate, since an unambiguous definition of
local quantities is not generally possible in general relativity. One
possible way to do so is to construct gauge invariant
quantities, together with a judicious choice of coordinates to specify
the points at which these quantities are evaluated.  Two useful
scalars can be constructed via the Riemann tensor.  These are the
Kretschmann $K_1\equiv R_{ABCD} R^{ABCD}$ and the Chern-Pontryagin
$K_2\equiv{}^*R_{ABCD} R^{ABCD}$ scalars. Related to these, the
principal invariants of the Weyl tensor are $I_1\equiv C_{ABCD}
C^{ABCD}$ and $I_2\equiv {}^*C_{ABCD} C^{ABCD}$. In addition,
Newman-Penrose scalars are useful to describe particular
characteristics of the solution. 

Here, we evaluate these quantities in the case of a single vortex on
the boundary.  Since the vortices we have seen are approximately
axially symmetric, it is convenient to introduce boundary coordinates
$(t,\varrho,\phi)$. The boundary functions then take the form
$\rho=\rho(\varrho), u_{\varrho}=0,u_{\phi}=u_{\phi}(\varrho)$, where
we set $\varrho=0$ at the center of the vortex. With this assumption, the  zeroth
order bulk metric \eqref{eq:metric0} takes the form,
\begin{equation}\label{eq:metric1}
  ds^2_{[0]} = -2 u_{\mu} dx^{\mu} dr + \left( r^2 P_{\mu\nu} - F(r,\varrho) u_{\mu} u_{\nu} \right) dx^\mu dx^\nu.
\end{equation}
Here, $F(r,\varrho) = \frac{\rho(\varrho)}{r}$, and the horizon is
located at $r=\rho(\varrho)$ to zeroth order.

We also introduce the following null tetrad:
\begin{align}
l^A&=\partial_r^A,\\ n^A &= u_0 \partial_t^A - \frac{1}{2}(r^2 - F(r,\varrho)) \partial_r^A - \frac{u_{\phi}}{\varrho^{2}} \partial_{\phi}^A, \\
m^A &= \frac{u_{\phi}}{\sqrt{2}r\varrho} \partial_t^A + i \frac{1}{\sqrt{2}r} \partial_{\varrho}^A - \frac{u_0}{\sqrt{2}r\varrho} \partial_{\phi}^A  ,
\end{align}
which satisfies $ - l^A n_A = m^A \bar m_A = 1$. With the tetrad, we construct the Newman-Penrose scalars,
$\Psi_0 = C_{ABCD} l^A m^B l^C m^D$, 
$\Psi_1 = C_{ABCD} l^A n^B l^C m^D$, 
$\Psi_2 = C_{ABCD} l^A m^B \bar m^C n^D$, 
$\Psi_3 = C_{ABCD} l^A n^B \bar m^C n^D$, 
$\Psi_4 = C_{ABCD} n^A \bar m^B n^C \bar m^C$, which  are related to the mass aspect, angular momentum and radiation
in the system.

We evaluate all of these quantities for our vortex solution to leading
order, and in the non-relativistic case,
\begin{eqnarray}
K_1 &\sim& 12\left(2+\frac{\rho(\varrho)^{2}}{r^6}\right) \, \, , \, \, K_2 \sim  36 \frac{\rho(\varrho)^{2}}{r^7}\omega \, ,\\
\label{eq:weylinvariants}I_1 &\sim& 12 \frac{\rho(\varrho)^{2}}{r^6} \, \, , \, \, I_2 \sim 36 \frac{\rho(\varrho)^{2}}{r^7}\omega \, , \\
\Psi_0 = 0 \, \, , \, \, \Psi_1 \sim   \frac{\sqrt{2}}{8 \varrho r^3} \omega\, \, , \, \,
\Psi_2 &\sim& -\frac{\rho(\varrho)}{2 r^3} - i \frac{3 \rho(\varrho)}{4 r^4} \omega \, \, , \, \,
\Psi_3 \sim -\frac{\sqrt{2}(r^3 + \rho(\varrho))}{8\varrho r^4} \omega \, \, , \, \,
\Psi_4 \sim - i \frac{3\rho(\varrho)}{4 r^2 \varrho^{2}} (\varrho^{2}\omega - 2 u_{\phi}) \, .
\end{eqnarray}
(evaluation of the horizon is obtained by $r\rightarrow\rho(\varrho)^{1/3}$).
Notice that $\Psi_4$ is purely imaginary, which implies a single polarization of
gravitational waves due to the axisymmetric structure of the vortex. 

\begin{figure}
  \centering
  \includegraphics[height=2in,width=2.2in]{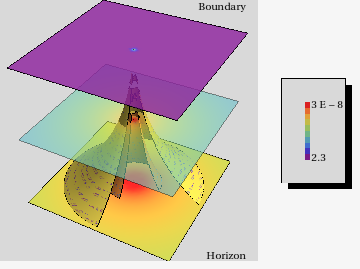}
  \includegraphics[height=2in,width=2.2in]{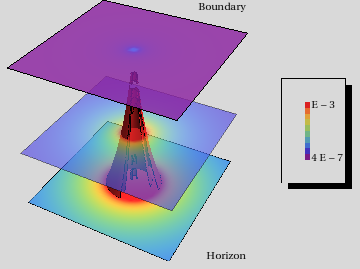}
  \includegraphics[height=2in,width=1.7in]{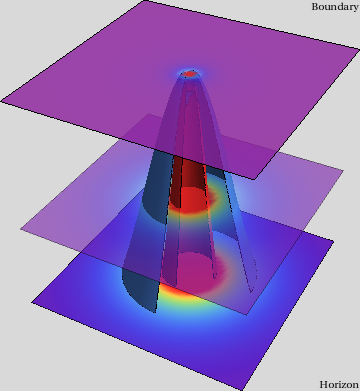}
  \caption{Contour plots for $\Psi_1$, $\Im(\Psi_2)$ and $\Psi_3$
    (left to right) illustrating the vortex structure from the horizon
    to the boundary. Note that the ``conical'' structure is a result
    of the dependence upon the radial coordinate of the plotted
    quantities (see Eq.~\eqref{eq:weylinvariants}). (In contrast to
    Fig.~\ref{fig:bulkC2}, we are not rescaling here by any powers of
    $r$.)}
  \label{fig:Psis}
\end{figure}
\begin{figure}
  \centering
  \includegraphics[width=2.5in,width=2.5in]{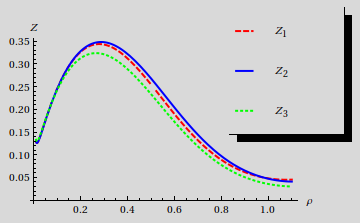}
  \caption{Radial profile of the ``enstrophy'' at the horizon, for a
    vortex solution, as calculated by Eq.~\eqref{enstrophy-r}, the
    square of the Pontryagin density~\cite{Adams:2013vsa}, and
    $[\Im(\Psi_2)]^2$~\cite{Eling:2013sna} (labelled $Z_1,Z_2$ and
    $Z_3$, respectively).  Each curve has been rescaled by a trivial
    constant factor for easier comparison.}
  \label{fig:Zcomps}
\end{figure}

Notice that most of these curvature quantities depend to leading order
on the vorticity. In fact, they are related through the identity $I_1
- i \, I_2 = 16 (3\, \Psi_2^2 + \Psi_0 \Psi_4 - 4 \Psi_1 \Psi_3)$.
Thus, there is some freedom in choosing which curvature quantities to
use to analyze the structure which arises in the
bulk. Fig.~\ref{fig:Psis} displays the behavior of $\Psi_1,
\Im(\Psi_2)$ and $\Psi_3$ as functions of $(r,\varrho,\phi)$ at fixed
$t$, while Fig.~\ref{fig:bulkC2} illustrates the behavior of $I_1$ and
$I_2$ for one of the solutions we obtained in our simulations.  Recent
works have proposed several different curvature quantities to
represent the enstrophy. For instance, at the horizon, the squares of $K_2$ and
$\Im(\Psi_2)$, respectively, have been suggested in \cite{Adams:2013vsa} and
\cite{Eling:2013sna}.  These quantities can be extended
throughout the bulk and compared with the quantity introduced in
Appendix~\ref{app:Z}. Fig.~\ref{fig:Zcomps} illustrates the radial
profile of the three quantities, showing good agreement.

\bibliography{fgbib}

\end{document}